\documentstyle[12pt,epsfig]{article}
\newcommand{\bea}{\begin{equation}}
\newcommand{\eea}{\end{equation}}
\newcommand{\beq}{\begin{eqnarray}}
\newcommand{\eeq}{\end{eqnarray}}
\newcommand{\ba}{\begin{array}}
\newcommand{\ea}{\end{array}}
\newcommand{\dd}{{\rm d}}
\newcommand{\dfrac}{\displaystyle\frac}
\newcommand{\nn}{\nonumber}
\newcommand{\gmu}{\gamma_\mu}
\newcommand{\gmup}{\gamma^\mu}
\newcommand{\gfi}{\gamma_5}

\newcommand{\smunu}{\sigma_{\mu\nu}}
\newcommand{\smunup}{\sigma^{\mu\nu}}
\newcommand{\eps}{\epsilon}

\def\ijmp#1#2#3{{\it Int. Jour. Mod. Phys. }{\bf #1~}(19#2)~#3}
\def\fp#1#2#3{{\it Fortschr. Phys. }{\bf #1~}(19#2)~#3}
\def\plb#1#2#3{{\it Phys. Lett. }{\bf B#1~}(19#2)~#3}
\def\zpc#1#2#3{{\it Z. Phys. }{\bf C#1~}(19#2)~#3}
\def\prl#1#2#3{{\it Phys. Rev. Lett. }{\bf #1~}(19#2)~#3}
\def\rmp#1#2#3{{\it Rev. Mod. Phys. }{\bf #1~}(19#2)~#3}
\def\prep#1#2#3{{\it Phys. Rep. }{\bf #1~}(19#2)~#3}
\def\prd#1#2#3{{\it Phys. Rev. }{\bf D#1~}(19#2)~#3}
\def\npb#1#2#3{{\it Nucl. Phys. }{\bf B#1~}(19#2)~#3}
\def\mpl#1#2#3{{\it Mod. Phys. Lett. }{\bf #1~}(19#2)~#3}
\def\arnps#1#2#3{{\it Annu. Rev. Nucl. Part. Sci. }{\bf #1~}(19#2)~#3}
\def\sjnp#1#2#3{{\it Sov. J. Nucl. Phys. }{\bf #1~}(19#2)~#3}

\def\ptp#1#2#3{{\it Prog. Theor. Phys. }{\bf #1~}(19#2)~#3}
\newcommand{\gsim}{\raisebox{-0.13cm}{~\shortstack{$>$ \\[-0.07cm] $\sim$}}~}
\def\pplus{{\mathbf{\hat p}_{\mathbf{+}}}}

\def\kplus{{\mathbf{\hat k}_{\mathbf{+}}}}

\def\qplus{{\mathbf{q}_{\mathbf{+}}}}
\def\qminus{{\mathbf{q}_{\mathbf{-}}}}

\def\sone{{\mathbf{s}_{\mathbf{1}}}}
\def\stwo{{\mathbf{s}_{\mathbf{2}}}}
\def\soner{{\mathbf{s}^{\mathbf{*}}_{\mathbf{1}}}}
\def\stwor{{\mathbf{s}^{\mathbf{*}}_{\mathbf{2}}}}

\def\ppmn{{\mathbf{p}_{\mathbf{\pm}}}}
\def\pmpn{{\mathbf{p}_{\mathbf{\mp}}}}
\def\kplusn{{\mathbf{k}_{\mathbf{+}}}}

\def\kpmn{{\mathbf{k}_{\mathbf{\pm}}}}
\def\kmpn{{\mathbf{k}_{\mathbf{\mp}}}}
\def\qplusn{{\mathbf{\hat q}_{\mathbf{+}}}}
\def\qminusn{{\mathbf{\hat q}_{\mathbf{-}}}}
\def\qpmn{{\mathbf{q}_{\mathbf{\pm}}}}
\def\qmpn{{\mathbf{q}_{\mathbf{\mp}}}}
\begin{document}

\thispagestyle{empty}

\begin{flushleft}
DESY 98--195
\\
KA--TP--20--1998
\\
{\tt hep-ph/9812298}
\\
October 2001 (rev.)
\end{flushleft}

\begin{center}

{\Large \bf
Top Dipole Form Factors and Loop--induced 
\\
\vspace{2mm}
CP violation in Supersymmetry
}

\vspace{5mm}

{\large W. Hollik$^{a,b}$, 
	J.I. Illana$^c$, 
	S. Rigolin$^d$}
	
{\large C. Schappacher$^a$, 
	D. St\"ockinger$^a$}

\vspace{5mm}

{\sl $^a$ Institut f\"ur Theoretische Physik, 
	      Universit\"at Karlsruhe,\\
	      D--76128 Karlsruhe, Germany}

{\sl $^b$ Theoretical Physics Division, CERN,\\
              CH--1211 Geneva 23, Switzerland}

{\sl $^c$ Deutsches Elektronen--Synchrotron DESY,\\
	      D--15738 Zeuthen, Germany}

{\sl $^d$ Departamento de F{\'\i}sica Te\'orica, 
	      Universidad Aut\'onoma de Madrid,\\ 
	      Cantoblanco, E--28049 Madrid, Spain} 	       

\end{center}


\begin{abstract}

The one--loop Minimal Supersymmetric Standard Model (MSSM) contributions to the 
weak and electromagnetic dipole form factors of the top quark are presented. 
Far from the $Z$ peak, they are not sufficient to account for all the new 
physics effects.
In the context of the calculation of the process $e^+e^-\to t\bar{t}$ 
to one loop in the MSSM, we compare the impact on the phenomenology of the 
CP--violating dipole form factors of the top quark with the contribution from 
CP--violating box graphs. 
Some exemplificative observables are analyzed and the relevance of both the 
contributions is pointed out. 
The one--loop expressions for the electromagnetic and weak dipole 
form factors in a general renormalizable theory and the SM and MSSM couplings 
and conventions are also given.

\end{abstract}
\vfill
\clearpage

\section{Introduction}

The investigation of the electric and magnetic dipole moments of fermions 
provides deep insight in particle theory. The measurement of the 
intrinsic {\em magnetic dipole moment} (MDM) of the electron 
proved the correctness of the hypothesis of half--integer--spin particles 
\cite{gelectron} and is one of the most spectacular achievements of 
quantum field theory predictions. More precise studies of electron 
and muon showed afterwards the presence of {\em anomalous} contributions to 
the MDM (AMDM). Their predictions constitute one of the most spectacular
achievements of Quantum Field Theory and imply very accurate tests of the 
quantum structure of the Standard Model (SM). 
The measurements of the $(g_e-2)$ and $(g_\mu-2)$ 
available \cite{databook} are in perfect agreement with the  
SM predictions to several orders in the perturbative expansion of the theory
(cf. \cite{smemdm} and references therein). Furthermore, 
with the expected precision at the E821 Brookhaven experiment \cite{brook} it 
will be possible to improve the previous measurement of $(g_\mu-2)$ by a 
factor 20. Therefore the MDMs can be used, together with the precision tests at 
the $Z$ resonance from LEP and SLC and the new results of LEP2 and TEVATRON, to 
set bounds on possible new physics effects beyond the SM \cite{munpemdm}.  

The importance of the analysis of the {\em electric dipole moment} (EDM) of 
elementary and composite particles is intimately related to the CP violating 
character of the theory. In the electroweak SM there is only one 
possible source of CP violation, the $\delta_{\rm CKM}$ phase of 
the Cabibbo--Kobayashi--Maskawa (CKM) mixing matrix 
for quarks \cite{ckm}. Currently the only place where CP violation has been 
measured, the neutral $K$ system, fixes the value of this phase but does not 
constitute itself a test for the origin of CP violation \cite{cpreview}. 
On the other hand, if the baryon asymmetry of the universe has been 
dynamically generated, CP must be violated. The SM cannot account for the 
size of the observed asymmetry \cite{b-asym}. In extended models (beyond the 
SM) many other possible explanations of CP violation can be given. In 
particular, in supersymmetric (SUSY) models \cite{susycp,dugan} CP violation 
can appear assuming complex soft--SUSY--breaking terms. 
Two physical phases remain in the GUT constrained MSSM 
\cite{dugan,gavela,relax}, enough to provide the correct size of baryon 
asymmetry in some range of parameters \cite{b-susyasym}.
But the most significant effect of the CP violating phases in the phenomenology 
is their contribution to the EDMs \cite{dipoles}. Unlike the SM, where the 
contribution to the EDM of fermions arises beyond two loops \cite{smedm}, the 
MSSM can give a contribution already at the one--loop level \cite{susycp}. 

%
The measurements of the neutron, electron and muon EDMs \cite{edm-n,edm-l} 
constrain the phases and the supersymmetric spectrum in a way that {\em may} 
demand fine tuning (supersymmetric CP problem): either the SUSY particles very 
heavy (several TeV \cite{nath1}) or the phases are of ${\cal O}(10^{-2})$ 
\cite{susycp}. 
Very large soft--SUSY--breaking masses are unappealing as it seems natural to 
demand the SUSY spectrum to be at the electroweak scale.\footnote{Moreover
if the SUSY spectrum is in the TeV region this could also give rise to relic 
densities unacceptably large.} 
%
%
On the other side, the experimental constraints can be met taking general 
universal soft--SUSY--breaking terms and vanishing SUSY CP phases. 
In this case the CP violation is originated via the usual SM CKM mechanism 
and the supersymmetric spectrum affects the observables only through 
radiative contributions. 
In this scenario it is difficult to contruct models in which the SUSY 
phases naturally vanish \cite{phase0} and at the same time provide some 
other non--standard mechanism for explaining electroweak baryogenesis. 
%
%
There exist also ways of naturally obtaining small non--zero SUSY CP phases 
which leave sufficient CP violation for baryogenesis \cite{relax}. 
But, in general, they do not lead to observable differences from
the SM. In models with potentially observable predictions one has to
relax the assumption of soft--term universality.
%
%
Several attempts have been made, following this 
direction, to use CP violation from top--squark mixing: a complex 
parameter $A_t$ would yield large CP violating effects in collider 
processes involving top quarks \cite{cptop,aoki-oshimo}.\footnote{ 
Large non--SM CP violating top--quark couplings could be probed at high
energy colliders like the NLC \cite{nlc}.} 
%
%
Besides, due to renormalization--group relations, the phase of $A_t$ is 
constrained by the EDM of the neutron \cite{garisto}. 
%
%
One can satisfy the experimental constraints due to cancellation among the 
different components of the neutron EDM (constituent quarks and gluons), 
the SUSY phases can still be kept of ${\cal O}(1)$ and the SUSY spectrum at 
the electroweak scale satisfying the experimental bounds \cite{nath2}.
%
%
%
In \cite{falkolive,kane} it is shown that large CP violating phases in
the MSSM are compatible with the bounds on 
the electron and neutron EDMs as well as with the cosmological relic 
densities.
%
%
In view of all these arguments we keep our analysis completely general and 
consider the SUSY CP--phases as free parameters.

Some attention has also been payed to the study of possible CP violating 
effects in the context of R--parity violating models \cite{abel}. In this 
class of models new interactions appear providing extra sources of CP 
violation (still preventing fast proton decay). They can explain the 
CP violation in the $K$ system (with no need of the CKM phase) without 
introducing anomalous Flavor Changing Neutral Current (FCNC) contributions 
\cite{fcnc-cp}. 
In the following we restrict our discussion to the simpler 
case of models with conserved R-parity.

In addition a certain amount of investigation has been devoted to the analysis 
of {\em weak dipole moments} (WDM). The WDMs are defined, in analogy to the 
usual DMs, taking the corresponding on--shell chirality--flipping form factors 
of the $Zff$ effective vertex. 
The SM one--loop contribution to 
the {\em anomalous weak magnetic dipole moment} (AWMDM) has been calculated 
for the $\tau$ lepton and the $b$ quark in \cite{ber95,ber97}. 
The CP--violating {\em weak electric dipole moment} (WEDM) is in the SM 
a tiny three--loop effect. The WDMs are gauge invariant and can directly 
connected to physical observables. While for the $\tau$ case, using 
appropriate observables \cite{ber95,ber94,ber95p}, an experimental analysis 
is feasible, for the $b$ case the situation is complicated by hadronization 
effects \cite{mele}.
The SM predictions are far below the sensitivity reachable at LEP 
\cite{ber95} but non--standard interactions can enhance these
expectations (2HDM \cite{ber95b}, MSSM \cite{hirs1}) especially for the 
(CP--violating) WEDMs (2HDM \cite{2hdm-cp}, leptoquark models 
\cite{bernre97b}, MSSM \cite{hirs2}).
The experimental detection of non--zero AWMDM or WEDM of heavy fermions, at 
the current sensitivity, would be a clear evidence of new physics beyond the SM.

In the perspective of the next generation of linear $e^+ e^-$ colliders, 
we extend the previous analyses on the WDMs to consider the $t$ quark
{\em dipole form factors} (DFF) and, in particular, the 
CP--violating ones. In \cite{vienna} an independent analysis of the $t$ quark 
EDFF and WEDFF can be found.
%
Since the $t$ quark is very heavy one expects this fermion to be the best 
candidate to have larger DFFs. Beyond the $Z$ peak 
($s>4m^2_t$) other effects are expected to give contributions to 
the physical observables. In fact the DFFs in a general model are not 
guaranteed to be gauge independent. An exception is the 
one--loop MSSM contribution to the CP--violating {\em electromagnetic} and 
{\em weak} DFFs (EDFF and WEDFF) which do not involve any gauge boson in 
internal lines. Anyway, although this contribution is indeed gauge invariant 
any CP--odd observable will be sensitive to not only the DFFs but to the 
complete set of one--loop CP--violating diagrams involved. In this work we 
present the full calculation of the expectation value of a set of these 
observables in the context of the MSSM and compare the size of the different 
contributions.
 
The paper is organized as follows.
In Section 2 we present the general effective vertex describing the interaction
of on--shell fermions with a neutral vector boson. The definitions and the 
generic expressions of the DFFs for all the contributing topologies at the 
one--loop level are given.
In Section 3 we briefly discuss the cancellation of the dipoles in the
supersymmetric limit.
The numerical results for the $t$ quark EDFFs and WEDFFs at $\sqrt{s}=500$ GeV 
are presented in Section 4. 
In Section 5 we evaluate specific CP--odd observables for the $t$ quark pair
production in $e^+e^-$ colliders to one loop in the MSSM 
and compare the influence of the $t$ EDFF and WEDFF with
that of the CP--violating box diagrams.
Our conclusions are presented in Section 6.
In the Appendices one can find the definition of the one--loop 3--point 
tensor integrals employed in the expressions for the dipoles as well as
the SM and MSSM couplings and conventions that have been used. 


%
\section{The dipole form factors}

\subsection{The $Vff$ effective vertex}
%

The most general effective Lagrangian describing the interaction 
of a neutral vector boson $V$ with two fermions can be written, using at 
most dimension five operators, as a function of ten independent terms:
\beq
\hspace{-2mm}
{\cal L}_{Vff} 
&=& 
V^\mu (x) \bar{\Psi} (x) \Bigg[ \gmu \left(g_{\rm V} - g_{\rm A} \gfi \right) 
\ + \ {\rm i} \stackrel{\leftrightarrow}{\partial_\mu} 
\left(g_{\rm M} + {\rm i}g_{\rm E} \gfi \right)   \nn \\
& & \qquad \qquad \qquad +
{\rm i}\stackrel{\leftrightarrow}{\partial^\nu} \smunu 
\left( g_{\rm TS} + {\rm i} g_{\rm TP} \gfi \right) \Bigg] \Psi (x) \ \nn \\
&+& 
\left({\rm i} \partial^\mu V^\nu (x) \right) \bar{\Psi} (x) \Bigg[ 
g_{\mu \nu} \left({\rm i}g_{\rm S} + g_{\rm P} \gfi \right) \ + \ 
\smunu \left({\rm i} g_{\rm TM} + g_{\rm TE} \gfi \right) \Bigg] \Psi (x).
\label{efflag}
\eeq
The first two coefficients, i.e. $g_{\rm V}$ and $g_{\rm A}$, are the 
usual vector and axial--vector couplings. They are connected to chirality 
conserving dimension four operators. All the other coefficients in 
Eq.~(\ref{efflag}) multiply chirality flipping dimension five operators 
and can receive a contribution only through radiative corrections in a
renormalizable theory. The operators associated to $g_{\rm V}$, $g_{\rm A}$, 
$g_{\rm M}$, $g_{\rm P}$, $g_{\rm TM}$ and $g_{\rm TP}$ are even under a CP 
transformation. The presence of non vanishing $g_{\rm E}$, $g_{\rm S}$, 
$g_{\rm TE}$ and $g_{\rm TS}$ yields a contribution to CP--violating 
observables. 
In Table~\ref{tab11} we summarize the C, P, T and chirality properties of 
each operator introduced in the effective Lagrangian.
\begin{table}
\vspace{0.3cm}
\caption{\em C, P, T properties of the operators in the effective Lagrangian of 
Eq.~(\ref{efflag}). Their chirality flipping behavior is also displayed.}
\label{tab11}
\vspace*{0.2cm}
\begin{center}
\begin{tabular}{|l|c|c|c|c|c|}  
\hline 
Operator & Coefficient & P & CP & T & Chirality Flip \\
 \hline \hline
$V^\mu \bar{\Psi} \gmu \Psi$ 
         & $g_{\rm V}$ & $+$ & $+$ & $+$ & NO \\
$V^\mu \bar{\Psi} \gmu \gfi \Psi$ 
         & $g_{\rm A}$ & $-$ & $+$ & $+$ & NO \\
$V^\mu \bar{\Psi}{\rm i}\stackrel{\leftrightarrow}{\partial_\mu} \Psi$
         & $g_{\rm M}$ & $+$ & $+$ & $+$ & YES \\
$V^\mu \bar{\Psi} \stackrel{\leftrightarrow}{\partial_\mu} \gfi \Psi$
         & $g_{\rm E}$ & $-$ & $-$ & $-$ & YES \\
\hline 
$V^\mu \bar{\Psi}{\rm i}\stackrel{\leftrightarrow}{\partial^\nu} \smunu \Psi$
         & $g_{\rm TS}$ & $+$ & $-$ & $-$ & YES \\
$V^\mu \bar{\Psi} \stackrel{\leftrightarrow}{\partial^\nu} \smunu \gfi \Psi$
         & $g_{\rm TP}$ & $-$ & $+$ & $+$ & YES \\
$\left(\partial \cdot V \right) \bar{\Psi} \Psi$  
         & $g_{\rm S}$ & $+$ & $-$ & $-$ & YES \\
$\left({\rm i}\partial \cdot V \right) \bar{\Psi} \gfi \Psi$ 
         & $g_{\rm P}$ & $-$ & $+$ & $+$ & YES \\
$\left(\partial^\mu V^\nu \right) \bar{\Psi} \smunu \Psi$ 
         & $g_{\rm TM}$ & $+$ & $+$ & $+$ & YES \\
$\left({\rm i}\partial^\mu V^\nu \right) \bar{\Psi} \smunu \gfi \Psi$ 
         & $g_{\rm TE}$ & $-$ & $-$ & $-$ & YES \\
\hline
\end{tabular}
\end{center}
\vspace*{0.2cm}
\end{table}

By Fourier transform of Eq.~(\ref{efflag}) one obtains the most general 
Lorentz structure for the vertex $Vff$ in the momentum space:
\beq
\Gamma^{Vff}_\mu & = & {\rm i}  \Bigg[
\gmu \left(f_{\rm V} - f_{\rm A} \gfi \right) \ + \ 
(q-\bar{q})_\mu \left(f_{\rm M} + {\rm i}f_{\rm E} \gfi \right) + \ p_\mu 
\left({\rm i}f_{\rm S} + f_{\rm P} \gfi \right) \nn \\ 
& & \qquad  +
(q - \bar{q})^\nu \smunu 
\left( f_{\rm TS} + {\rm i}f_{\rm TP} \gfi \right) \ + \ 
p^\nu \smunu \left({\rm i}f_{\rm TM} + f_{\rm TE} \gfi 
\right) \Bigg],
\label{efflagm}
\eeq
where $q$ and $\bar{q}$ are the outgoing momenta of the fermions and 
$p = (q+\bar{q})$ is the total incoming momentum of the neutral boson $V$. 
The form factors $f_i$ are functions of the kinematical invariants. 
Actually they are more 
general than the coefficients $g_i$. In fact any operator of dimension higher 
than five added to the Lagrangian of Eq.~(\ref{efflag}) is related to a new 
coefficient $g_i$. But every new coefficient can contribute only to the ten 
independent form factors $f_i$ introduced in Eq.~(\ref{efflagm}). The 
parameters $g_i$ and the form factors $f_i$ can be complex in general. 
Their real parts account for dispersive effects (CPT--even) whereas 
their imaginary parts are related to absorptive contributions.

It is possible to lower the number of independent form factors in 
Eq.~(\ref{efflagm}) by imposing on--shell conditions on the 
fermionic and/or bosonic fields. For instance, in the case of on--shell 
fermions, making use of the Gordon identities:
\beq
2 m_f ~\bar{u}~ \gmup ~v & = & \Big\{ \bar{u}~(q - \bar{q})^\mu~v + 
\bar{u}~{\rm i}~(q + \bar{q})_\nu \smunup ~v \Big\}, \nn \\
2 m_f ~\bar{u}~ \gmup \gfi ~v & = & \Big\{ \bar{u}~(q + \bar{q})^\mu
\gfi~v + \bar{u}~{\rm i}~(q - \bar{q})_\nu \smunup \gfi ~v \Big\}, \nn \\
0 & = & \Big\{ \bar{u}~(q + \bar{q})^\mu ~v + 
\bar{u}~{\rm i}~(q - \bar{q})_\nu \smunup v \Big\}, \nn \\
0 & = & \Big\{ \bar{u}~(q - \bar{q})^\mu~\gfi ~v + 
\bar{u}~{\rm i}~(q + \bar{q})_\nu \smunup \gfi ~v \Big\}~,
\label{gordon}
\eeq
one can eliminate $f_{\rm TM}$, $f_{\rm TE}$, $f_{\rm TS}$ and $f_{\rm TP}$ 
from the effective Lagrangian. The number of relevant form factors can be 
further reduced taking also the boson $V$ on its mass shell. In this case the 
condition $p_\mu \eps^\mu=0$ automatically cancels all the contributions 
coming from $f_{\rm S}$ and $f_{\rm P}$. The same situation occurs for 
off--shell vector boson $V$ when, in the process $e^+e^- \to V^* \to f\bar{f}$, 
the electron mass is neglected. Therefore $f_{\rm S}$ and $f_{\rm P}$ will
be ignored in the following. With all these assumptions the $Vff$ effective 
vertex for on--shell fermions is conventionally written as: 
\bea
\hspace{-5mm}
\Gamma^{Vff}_\mu(s) = {\rm i}e
        \Bigg\{\gamma_\mu \left[V^V_f(s)-A^V_f(s) \gamma_5\right] + 
        \sigma_{\mu\nu} (q+\bar{q})^\nu \Bigg[ {\rm i}\frac{a^V_f(s)}{2m_f} - 
                \frac{d^V_f(s)}{e}\gamma_5 \Bigg]\Bigg\},
\label{vertex}
\eea
where $e$ and $m_f$ are respectively the electric unit charge and the mass
of the external fermion. The form factors in Eq.~(\ref{vertex}) depend only on 
$s$. As mentioned above $V^V_f(s)$ and $A^V_f(s)$ parameterize the vector and 
axial--vector current. They are 
connected to the chirality conserving CP--even sector. The form factors 
$a^{V}_f(s)$ and $d^{V}_f(s)$ are known respectively as anomalous magnetic 
dipole form factor (AMDFF) and electric dipole form factor (EDFF). They 
are both connected to chirality flipping operators. In a renormalizable 
theory they can receive contributions exclusively by quantum corrections. 
The EDFFs contribute to the CP--odd sector and constitute a source of 
CP--violation.

The {\em dipole moments} (DM) are defined taking the corresponding vector 
bosons on shell, $s=M^2_V$.
For $V=\gamma$ one gets the usual definitions of the photon {\em anomalous 
magnetic dipole moment} (AMDM) and {\em electric dipole moment} (EDM).
The definitions of electric charge, magnetic and electric 
dipole moments \cite{iz} consistent with our convention for the covariant 
derivative (\ref{covder}) are respectively:
\beq
      \mbox{charge}&\equiv& -e\ V^\gamma_f(0)=e\ Q_f\ , \\
      \mbox{MDM}   &\equiv& \frac{e}{2m_f}(V^\gamma_f(0)+a^\gamma_f(0))\ , \\
      \mbox{EDM}   &\equiv& d^\gamma_f(0)\ .
\eeq
Thus the AMDM of a fermion is $a^\gamma_f(0)=-Q_f(g_f-2)/2$ with 
$g_f$ being the gyromagnetic ratio. The axial-vector 
coupling $A^\gamma_f$ vanishes. For $V=Z$, the quantities $a^w_f\equiv 
a^Z_f(M^2_Z)$ and $d^w_f\equiv d^Z_f(M^2_Z)$ define the {\em anomalous 
weak--magnetic dipole moment} (AWMDM) and the {\em weak--electric dipole 
moment} (WEDM).\footnote{ 
Massless and neutral fermions may have magnetic moments. The usual 
parameterization in (\ref{vertex}) must be generalized by the replacement 
$e\ a^V_f(s)/2m_f\to\mu^V(s)$. The dipole moments of neutrinos are then
given by $\mu^V(M^2_V)$.} 

%
%
\subsection{One--loop generic expressions of the dipole form factors}
\begin{figure}[t]
\begin{center}
\begin{tabular}{c}
\epsfig{figure=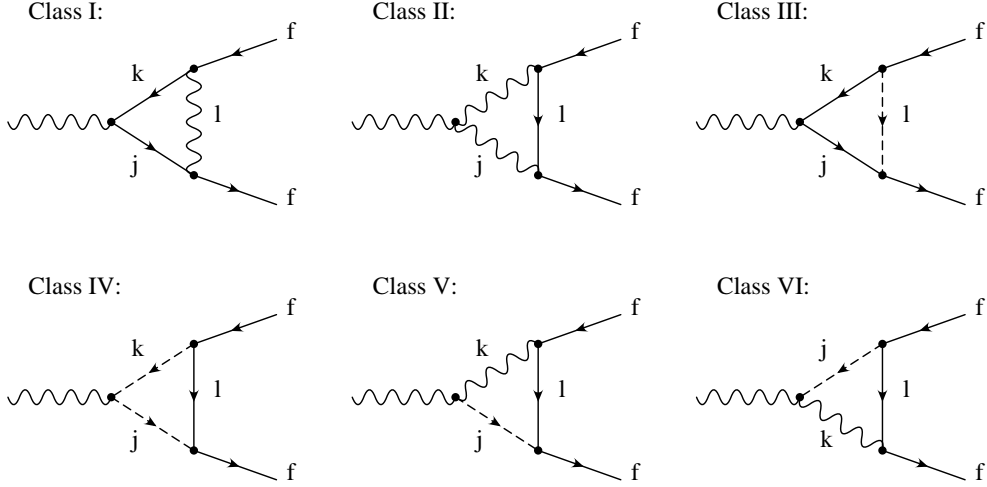,width=0.8\linewidth}
\end{tabular}
\end{center}
\caption{\em The one-loop $Vff$ diagrams with general couplings.
\label{fig1}}
\vspace*{0.2cm}
\end{figure}

%
%
All the possible one--loop contributions to the $a^V_f(s)$ and $d^V_f(s)$ 
form factors can be classified in terms of the six classes of triangle 
diagrams depicted in Fig.~\ref{fig1}. 
The vertices involved are labeled by generic couplings for vector bosons
$V^{(k)}_\mu=A_\mu,\ Z_\mu,\ W_\mu,\ W^\dagger_\mu$, fermions $\Psi_k$ and 
scalar bosons $\Phi_k$, according to the following interaction Lagrangian:
\beq
{\cal L} 
& = & 
     {\rm i} e J (W^\dagger_{\mu\nu}W^\mu V^\nu - W^{\mu\nu}W^\dagger_\mu V_\nu
             + V^{\mu\nu}W^\dagger_\mu W_\nu)
     + {\rm i} e G_{jk} V^\mu \Phi_j^\dagger\stackrel{\leftrightarrow}
             {\partial}_\mu\Phi_k \nn \\
& & 
     + \Big\{e \bar{\Psi}_f(S_{jk}-P_{jk}\gamma_5)\Psi_k\Phi_j\ 
     + e K_{jk} V^\mu V^{(k)}_\mu\Phi_j + {\rm h.c.} \Big\} \nn \\  
& &
     + e V^{(k)}_\mu\bar{\Psi}_j\gamma^\mu(V^{(k)}_{jl}-A^{(k)}_{jl}\gamma_5)
       \Psi_l \ .
\label{genlag}
\eeq
Every class of diagrams is calculated analytically and expressed in terms 
of the couplings introduced in (\ref{genlag}) and the one--loop 3--point 
integrals $\bar{C}_i$ (see App.~\ref{appendix-a}). 
The result is given in the 't Hooft-Feynman gauge.
\begin{itemize}
\item{[Class I]: vector--boson exchange:}
\beq
\frac{a^V_f(s)}{2 m_f}({\rm I}) 
&=& 
        \frac{\alpha}{4\pi}  \Big\{ 4m_f\sum_{jkl}{\rm Re} \Big[
   V^{(V)}_{jk}(V^{(l)}_{fj}V^{(l)*}_{fk} + A^{(l)}_{fj}A^{(l)*}_{fk})  \nn \\
& &
        \hspace{2truecm}+ A^{(V)}_{jk}(V^{(l)}_{fj}A^{(l)*}_{fk} + 
   A^{(l)}_{fj}V^{(l)*}_{fk}) \Big] \Big[2C^+_2 - 3C^+_1 + C_0\Big]_{kjl}
   \nn \\
& &
   \hspace{0.8truecm} + 4\sum_{jkl} m_k {\rm Re}
   \Big[V^{(V)}_{jk}(V^{(l)}_{fj}V^{(l)*}_{fk}-A^{(l)}_{fj}A^{(l)*}_{fk}) \nn \\
& & \hspace{2truecm}
    -A^{(V)}_{jk}(V^{(l)}_{fj}A^{(l)*}_{fk}-A^{(l)}_{fj}V^{(l)*}_{fk})\Big]
    \Big[2C^{+}_1 - C_0\Big]_{kjl} \Big\} 
\label{mdm1} \\
\frac{d^V_f(s)}{e}({\rm I}) 
&=& 
      \frac{\alpha}{4\pi}  \Big\{ 4m_f\sum_{jkl}{\rm Im}
      \Big[V^{(V)}_{jk}(V^{(l)}_{fj}A^{(l)*}_{fk}+A^{(l)}_{fj}V^{(l)*}_{fk}) 
      \nn \\
& &  \hspace{2truecm} 
      +A^{(V)}_{jk}(V^{(l)}_{fj}V^{(l)*}_{fk}+A^{(l)}_{fj}A^{(l)*}_{fk}) \Big]
      \Big[2C^{+-}_2-C^-_1 \Big]_{kjl}  \nn \\
& &
    \hspace{0.8truecm} -4\sum_{jkl}m_k{\rm Im}
    \Big[V^{(V)}_{jk}(V^{(l)}_{fj}A^{(l)*}_{fk}-A^{(l)}_{fj}V^{(l)*}_{fk})\nn \\
& & \hspace{2truecm}
        -A^{(V)}_{jk}(V^{(l)}_{fj}V^{(l)*}_{fk}-A^{(l)}_{fj}A^{(l)*}_{fk}) \Big]        
        \Big[2C^{+}_1-C_0 \Big]_{kjl} \Big\}
\label{edm1}
\eeq
In the case of gluon exchange one has to substitute $\alpha$ for $\alpha_s$
and $V^{(l)}_{fi}$ for $T_l$ (the SU(3) generators) and take
$A^{(l)}_{fj}=0$. The sum over the index $l$ yields a 
color factor $C_F=4/3$. The gluon contribution to the CP--violating form factor 
$d^V_f({\rm I}$) vanishes in general.
\item{[Class II]: fermion exchange and two internal vector bosons:}
\beq
\frac{a^V_f(s)}{2 m_f}({\rm II}) 
&=& 
       \frac{\alpha}{4\pi} \Big\{ 2m_f\sum_{jkl}{\rm Re}
   \Big[J (V^{(j)}_{fl}V^{(k)*}_{fl} + A^{(j)}_{fl}A^{(k)*}_{fl}) \Big] 
             \Big[4C^{+}_2 + C^+_1 \Big]_{kjl} \nn \\
& &
   \hspace{1truecm} - 6\sum_{jkl} m_l{\rm Re} 
   \Big[J(V^{(j)}_{fl}V^{(k)*}_{fl} - A^{(j)}_{fl}A^{(k)*}_{fl}) \Big]
             \Big[C^+_1 \Big]_{kjl} \Big\} \\
\label{mdm2}
\frac{d^V_f(s)}{e}({\rm II}) 
&=&
       -\frac{\alpha}{4\pi} \Big\{ 2m_f\sum_{jkl}{\rm Im}
   \Big[J(V^{(j)}_{fl}A^{(k)*}_{fl}+A^{(j)}_{fl}V^{(k)*}_{fl}) \Big]
            \Big[4C^{+-}_2-C^-_1 \Big]_{kjl} \nn \\
& &
  \hspace{1truecm} + 6\sum_{jkl} m_l{\rm Im}
   \Big[J(V^{(j)}_{fl}A^{(k)*}_{fl}-A^{(j)}_{fl}V^{(k)*}_{fl}) \Big]
           \Big[ C^+_1 \Big]_{kjl} \Big\}
\label{edm2}
\eeq
\item{[Class III]: scalar exchange:}
\beq
\frac{a^V_f(s)}{2 m_f}({\rm III}) &=&
       \frac{\alpha}{4\pi}  \Big\{2m_f\sum_{jkl}{\rm Re}
   \Big[V^{(V)}_{jk}(S_{lj}S^*_{lk} + P_{lj}P^*_{lk}) \nn \\
& &
        \hspace{2truecm} + A^{(V)}_{jk}(S_{lj}P^*_{lk} + P_{lj}S^*_{lk}) \Big] 
    \Big[2C^{+}_2 - C^+_1 \Big]_{kjl} \nn \\
& & \hspace{0.8truecm}
 - 2\sum_{jkl} {m}_k {\rm Re}\Big[V^{(V)}_{jk}(S_{lj}S^*_{lk} - P_{lj}P^*_{lk})
 \nn \\
& & \hspace{2truecm}
    -A^{(V)}_{jk}(S_{lj}P^*_{lk}-P_{lj}S^*_{lk}) \Big] 
    \Big[C^+_1 + C^-_1 \Big]_{kjl} \Big\} 
\label{mdm3}\\
\frac{d^V_f(s)}{e}({\rm III})
&=&  
         \frac{\alpha}{4\pi} \Big\{-2m_f\sum_{jkl}{\rm Im}
   \Big[V^{(V)}_{jk}(P_{lj}S^*_{lk}+S_{lj}P^*_{lk}) \nn\\
& & \hspace{2truecm}
   +A^{(V)}_{jk}(S_{lj}S^*_{lk}+P_{lj}P^*_{lk}) \Big] 
   \Big[2C^{+-}_2-C^-_1 \Big]_{kjl} \nn \\
& & \hspace{0.8truecm}
   +2\sum_{jkl} {m}_k {\rm Im}\Big[V^{(V)}_{jk}(P_{lj}S^*_{lk}-S_{lj}P^*_{lk})
   \nn\\
& & \hspace{2truecm}
   +A^{(V)}_{jk}(S_{lj}S^*_{lk}-P_{lj}P^*_{lk}) \Big]
   \Big[C^+_1+C^-_1 \Big]_{kjl} \Big\}
\label{edm3}
\eeq
\item{[Class IV]: fermion exchange and two internal scalars:}
\beq
\frac{a^V_f(s)}{2 m_f}({\rm IV})
& = & 
       - \frac{\alpha}{4\pi}  \Big\{ 2 m_f\sum_{jkl}{\rm Re}
   \Big[G_{jk}(S_{jl}S^*_{kl} + P_{jl}P^*_{kl}) \Big]
   \Big[2C^{+}_2 - C^+_1 \Big]_{kjl} \nn \\
& & \hspace{0.8truecm}
   +\sum_{jkl} {m}_l {\rm Re}\Big[G_{jk}(S_{jl}S^*_{kl} - P_{jl}P^*_{kl}) \Big] 
   \Big[2C^+_1 - C_0 \Big]_{kjl} \Big\} 
\label{mdm4}\\
\frac{d^V_f(s)}{e}({\rm IV})
&=&
         \frac{\alpha}{4\pi}  \Big\{2m_f\sum_{jkl} {\rm Im}
  \Big[G_{jk}(S_{jl}P^*_{kl}+P_{jl}S^*_{kl})\Big]\Big[2C^{+-}_2-C^-_1\Big]_{kjl}  
   \nn \\  
& & \hspace{0.8truecm}
   -\sum_{jkl} {m}_l {\rm Im}\Big[G_{jk}(S_{jl}P^*_{kl} - P_{jl}S^*_{kl}) \Big]
   \Big[2C^+_1 - C_0 \Big]_{kjl} \Big\}
\label{edm4}
\eeq
\item{[Class V+VI]: fermion exchange, one vector and one
                    scalar internal boson:}
\beq
\frac{a^V_f(s)}{2 m_f}({\rm V+VI})
&=&
         \frac{\alpha}{4\pi} 2\sum_{jkl}{\rm Re} 
   \Big[K_{jk}(V^{(k)}_{fl}S_{jl}^*+A^{(k)}_{fl}P_{jl}^*) \Big]
   \Big[C^+_1+C^-_1 \Big]_{kjl} \\ 
\label{mdm56}
\frac{d^V_f(s)}{e}({\rm V+VI})
&=&
         -\frac{\alpha}{4\pi} 2\sum_{jkl}{\rm Im}
   \Big[K_{jk}(V^{(k)}_{fl}P_{jl}^*+A^{(k)}_{fl}S_{jl}^*) \Big]
   \Big[C^+_1+C^-_1 \Big]_{kjl}
\label{edm56}
\eeq
\end{itemize}
In Eqs.~(\ref{mdm1}--\ref{edm56}) the shorthand notation $[\bar{C}]_{kjl}$ 
stands for the 3--point tensor integrals $\bar{C}(-\bar{q},q,M_k,M_j,M_l)$. 
The integrals appearing in the previous Eqs. are UV and IR finite. 
All the expressions are proportional to some positive 
power of a fermion mass, either internal or external, consistently with 
the chirality flipping character of the dipole moments.\footnote{
This is as expected when applying the mass--insertion method: to induce a 
flip in the fermion chirality one introduces a mass in either one of the 
internal fermion lines, picking a mass term from the propagator, or in the 
external fermion lines, using the equations of motion for the free fermion.}
For class V and VI diagrams the mass dependence is hidden in the product 
of the Yukawa couplings $S_{ij}$ ($P_{ij}$) and the dimensionful 
parameter $K_{ij}$. Hence, the heaviest fermions are the most promising 
candidates to have larger DFFs. Eqs.(\ref{mdm1}--\ref{edm56}) also show 
that, in general, the DFFs for massless fermions are not vanishing 
but proportional to masses of fermions running in the loop. The SM 
cancellation of the massless neutrino DFFs is only due to the absence of
right--handed neutrinos. Finally, 
notice that all the contributions to the EDFFs are proportional to the 
imaginary part of certain combinations of couplings. A theory with real 
couplings has manifestly vanishing EDFFs.
%


\section{Cancellation of the dipole moments in the supersymmetric limit} 

A general $Vff$ interaction is restricted to the form
(\ref{efflagm}) by Lorentz invariance. Since the Lorentz algebra is a
subalgebra of supersymmetry, this interaction is even more
constrained in a theory with unbroken supersymmetry. In \cite{SumRules} 
supersymmetric sum rules are derived that relate the
electric and magnetic multipole moments of any irreducible $N=1$
supermultiplet. Applied to the $Vff$ interaction between a
vector boson coupling to a conserved current and the fermionic
component of a chiral multiplet, these sum rules force the
gyromagnetic ratio to be $g_f=2$ and forbid an electric dipole
moment:
\begin{equation}
   a^V_f=d^V_f=0.
\end{equation}
The Lagrangian of the MSSM is supersymmetric when the soft--breaking terms
are removed. For non zero value of the $\mu$ parameter the Higgs potential 
has only a trivial minimum. Therefore to keep the particles massive and 
supersymmetry preserved at the same time the choice $\mu=0$ is necessary.
Then the Higgs potential is positive semi--definite and 
it has degenerate minima corresponding to $v\equiv v_1=v_2$ ($\tan\beta=1$).
The value of $M_A=0$ follows from such a configuration.
Finally the value of the common $v$ is fixed by the muon decay constant. 

In this supersymmetric limit the above mentioned sum rules are valid
and the magnetic and electric dipole form factors have to cancel. To verify
our expressions we checked this for the AWMDM of the $b$ quark \cite{hirs2}. 
Choosing the parameters $A_b=A_t=M_2=M_3=0$, $\mu=0$, $\tan\beta=1$ and
$M_A=0$, the SM gauge boson contribution to $a^w_b$
\cite{ber97} is indeed cancelled by the MSSM correction including
the two Higgs doublets: the gluon and gluino contribution
cancel among themselves and the neutralinos and charginos cancel the
gauge boson and Higgs contributions. A similar check has been performed
for the electric and weak--electric dipole form factors of the $t$ quark.

\section{SM and MSSM predictions for the top quark dipole form factors}

\subsection{SM}

The electroweak contributions to the magnetic and weak--magnetic dipole form 
factors for off--shell gauge bosons are gauge dependent. The pinch technique 
\cite{cornwall85} could be used to construct gauge--parameter independent 
magnetic dipoles in the class of $R_\xi$ gauges \cite{papa94,ber95b} but the 
prescription is not unique \cite{georg} and these quantities cannot be 
observable by themselves.
The QCD contributions (gluon exchange) to the $t$ (W)MDFF are gauge independent
and comparable in size to the electroweak predictions at $\sqrt{s}=500$ GeV 
in the 't Hooft--Feynman gauge \cite{ber95b} due to the large mass of the $t$ 
quark.

There is no contribution to the electric and weak--electric dipole form factors
to one loop in the SM.

\subsection{MSSM}

The triangle diagrams with SUSY particles are gauge independent by themselves
(no gauge or Goldstone bosons involved) but the ones including Higgs scalars 
and gauge bosons in the loop are not sufficient to keep the gauge invariance in 
the case of the magnetic dipole form factor. 
The CP--violating dipole form factors (electric and weak--electric) for which 
the Higgs sector of the MSSM is irrelevant, on the other hand, can be regarded
as gauge independent quantities. These form factors have also been considered in
\cite{vienna} and the numerical results in are in agreement with the ones 
presented here.
Since the MSSM contains a CP--conserving Higgs sector\footnote{
A one--loop non vanishing contribution to the WEDM is possible in a
general 2HDM \cite{2hdm-cp}.} 
and the SM provides contributions to the (W)EDMs beyond two loops, the 
relevant diagrams are the genuine SUSY graphs of classes III and IV. 

We investigate the SUSY contributions to the $t$ electric and weak--electric 
dipole form factors performing a parameter scan for which a fixed value 
$\sqrt{s}=500$ GeV has been chosen.\footnote{
We use the running coupling constants evaluated at $\sqrt{s}=500$ GeV,
$\alpha_s=0.092$, $\alpha=1/126$.}
It is important to point out that the region of the supersymmetric parameter 
space that provides the maximal contributions varies with $s$ due 
to threshold effects.
We scan the mass parameters $M_2$ and $|\mu|$ in a broad range and the 
CP--violating phases: $\varphi_\mu$, $\varphi_{\tilde t}$ and 
$\varphi_{\tilde b}$ (see Figs.~\ref{fig:4.1} and \ref{fig:4.2}). 
The gluino mass ($m_{\tilde g}=M_3$) is given by the GUT constraint 
(\ref{gut}).
We adopt a fixed value for the common scalar quark mass $m_{\tilde q}=200$ GeV: 
this is a plausible intermediate value; larger values decrease the effects. 
Finally, the moduli of the off--diagonal terms 
in the $\tilde t$ and $\tilde b$ mass matrices are also chosen at fixed values 
$|m^t_{LR}|=|m^b_{LR}|=200$ GeV, to reduce the number of free parameters.
The results are expressed in $t$ magnetons $\mu_t\equiv e/2m_t=5.64\times
10^{-17}\ e$cm.

The MSSM contributions to the top quark EDFF and WEDFF are analyzed below.

\subsubsection{Neutralinos and ${\tilde t}$ scalar quarks}

\begin{figure}
\begin{center}
\begin{tabular}{cc}
{\small Re$\{d^\gamma_t[{\tilde \chi}^0]\}$}            &
{\small Im$\{d^\gamma_t[{\tilde \chi}^0]\}$}            \\
\epsfig{file=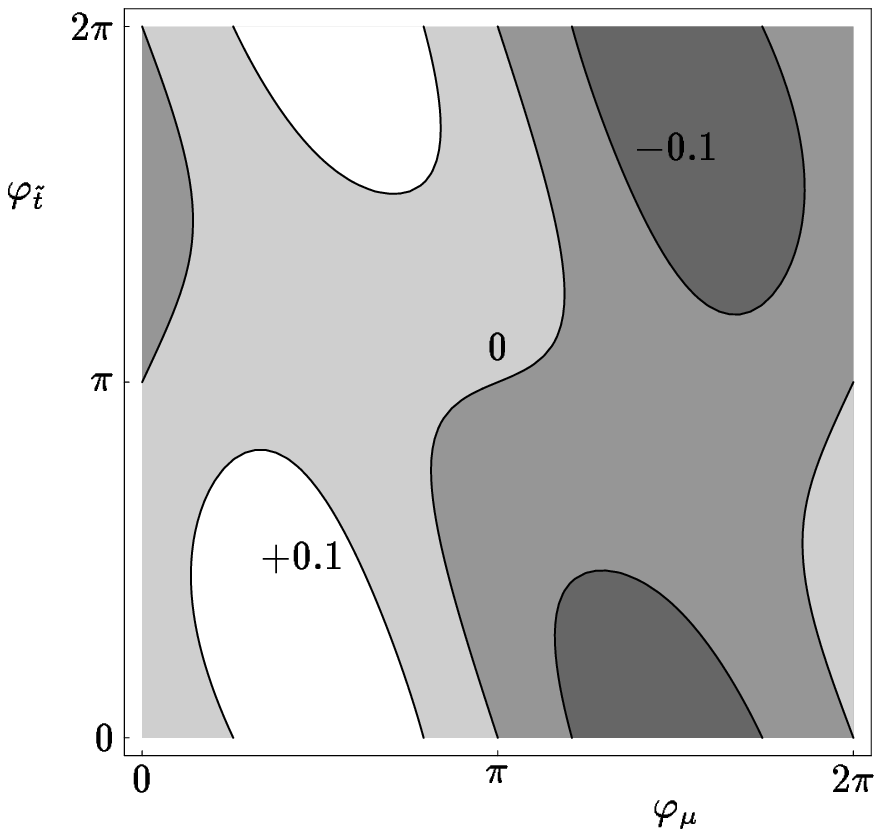,width=0.4\linewidth}        &    
\epsfig{file=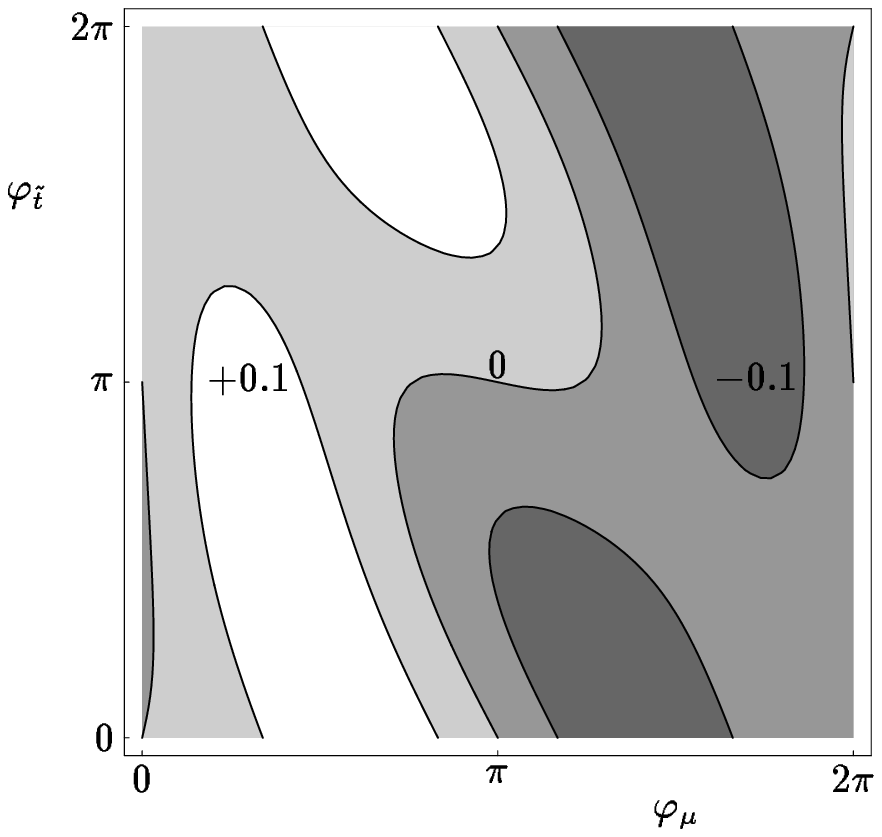,width=0.4\linewidth}       \\
{\small Re$\{d^Z_t[{\tilde \chi}^0]\}$}                 &
{\small Im$\{d^Z_t[{\tilde \chi}^0]\}$}                 \\
\epsfig{file=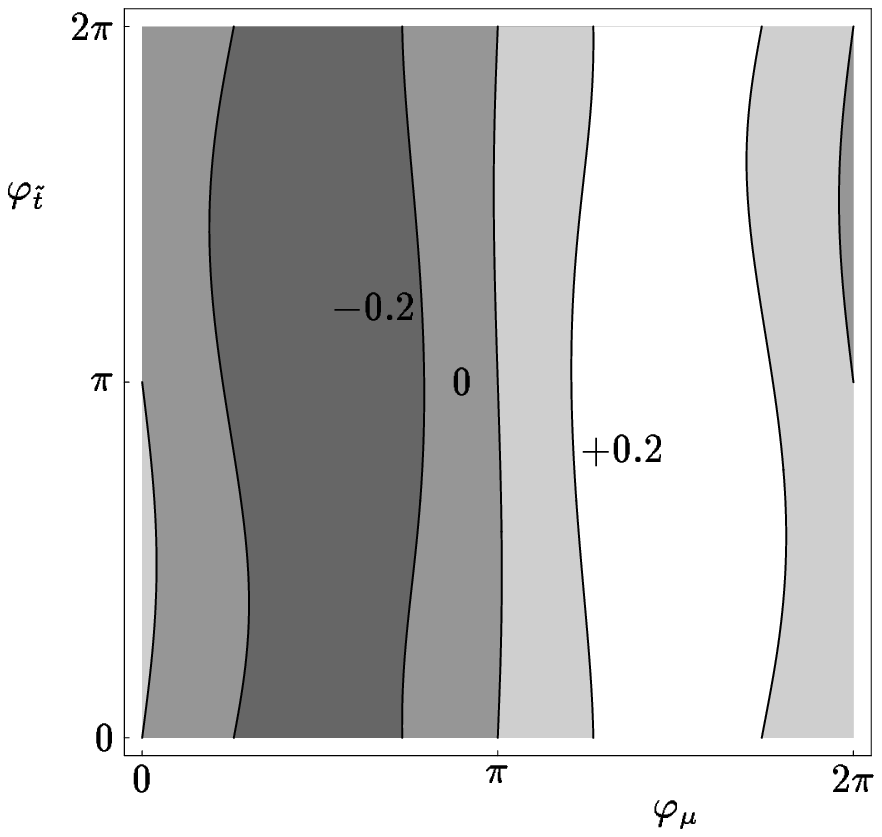,width=0.4\linewidth}       &    
\epsfig{file=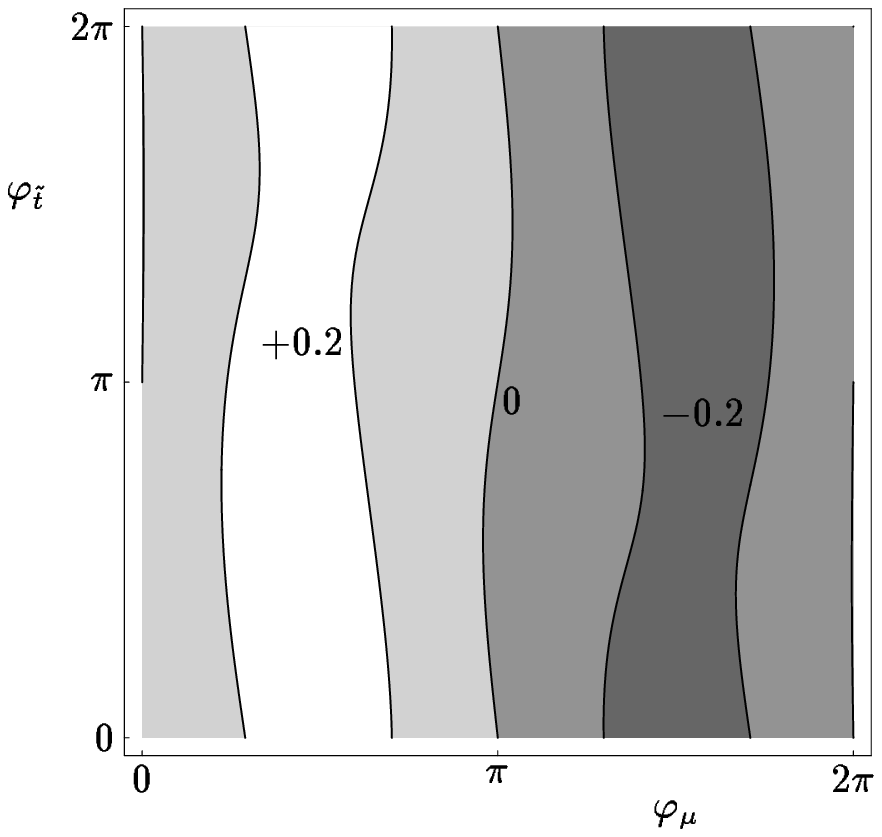,width=0.4\linewidth}      
\end{tabular}
\end{center}
\caption{\em Neutralino contribution to the real and imaginary parts of the $t$
         EDFF and WEFF [in $10^{-3}\mu_t$ units] in the plane $\varphi_{\tilde
         t}-\varphi_\mu$
         for $\tan\beta=1.6$ and the reference values $M_2=|\mu|=|m^t_{LR}|=
         200$ GeV at $\sqrt{s}=500$ GeV. \label{fig:4.1}}
\vspace*{0.3cm}
\end{figure}

They provide typically small contributions but quite sensitive to the
value of both the phases involved, $\varphi_\mu$ and $\varphi_{\tilde t}$
(Fig.~\ref{fig:4.1}). 

The results are larger for low $\tan\beta$ since the chirality flipping mass
terms are dominated by the $t$ quark, yielding a term proportional 
to $m_t\cot\beta$. A term proportional to the neutralino masses is also
present as well as a negligible one proportional to $m_b\tan\beta$.
The contributing diagrams belong to the classes III and IV for the $Z$ case
and only to class IV for the $\gamma$ case, as the neutralinos do not couple to
photons.
 
As a reference we take the representative values $M_2=|\mu|=200$ GeV and
$\varphi_\mu=-\varphi_{\tilde t}=\pi/2$.
For these inputs the results are
\beq
d^\gamma_t[{\tilde \chi}^0]     &=& (\ 0.080+0.081\ {\rm i})\times 10^{-3}\
\mu_t\ ,\\
d^Z_t[{\tilde \chi}^0]          &=& (-0.324+0.223\ {\rm i})\times 10^{-3} \
\mu_t\ .      
\eeq

\subsubsection{Charginos and ${\tilde b}$ scalar quarks}

\begin{figure}
\begin{center}
\begin{tabular}{cc}
{\small Re$\{d^\gamma_t[{\tilde \chi}^\pm]\}$}          &
{\small Im$\{d^\gamma_t[{\tilde \chi}^\pm]\}$}          \\
\epsfig{file=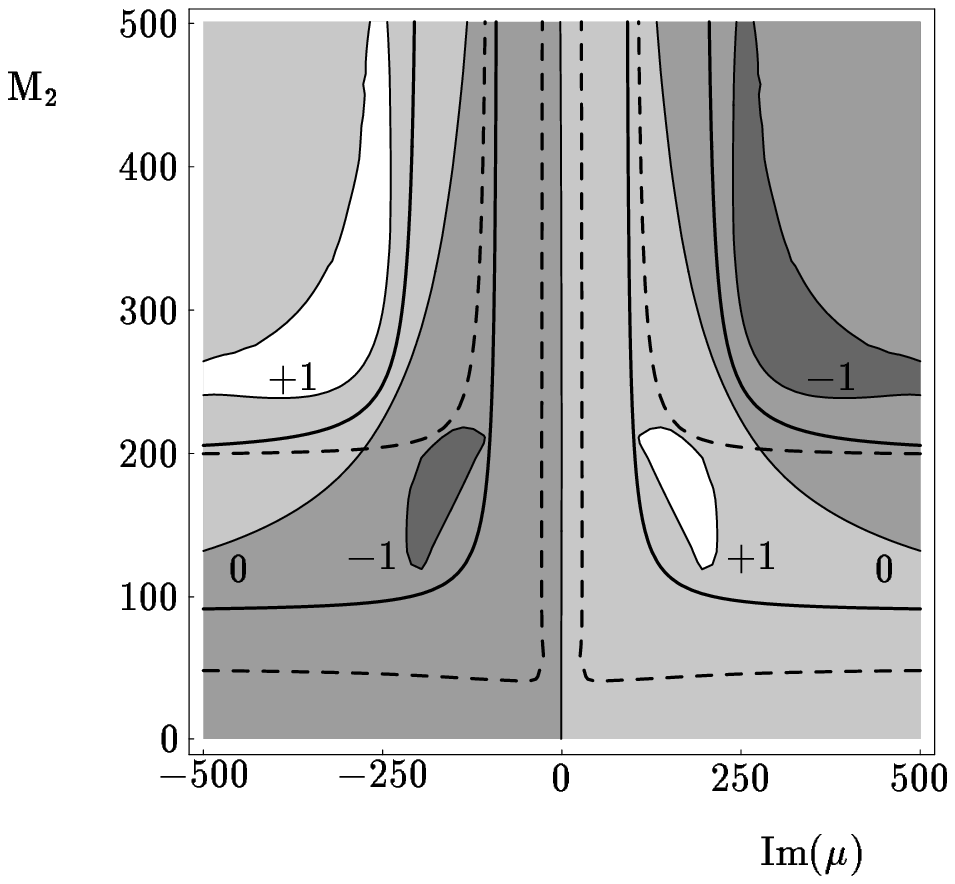,width=0.45\linewidth}     &    
\epsfig{file=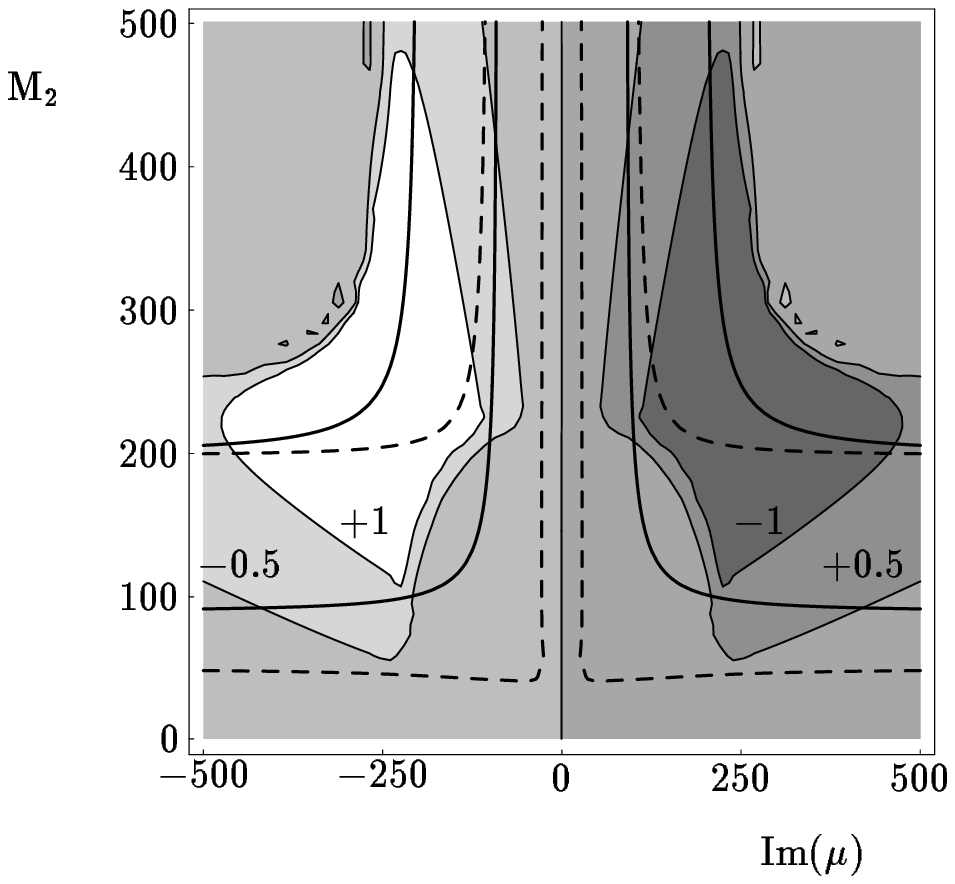,width=0.45\linewidth}     \\
{\small Re$\{d^Z_t[{\tilde \chi}^\pm]\}$}               &
{\small Im$\{d^Z_t[{\tilde \chi}^\pm]\}$}               \\
\epsfig{file=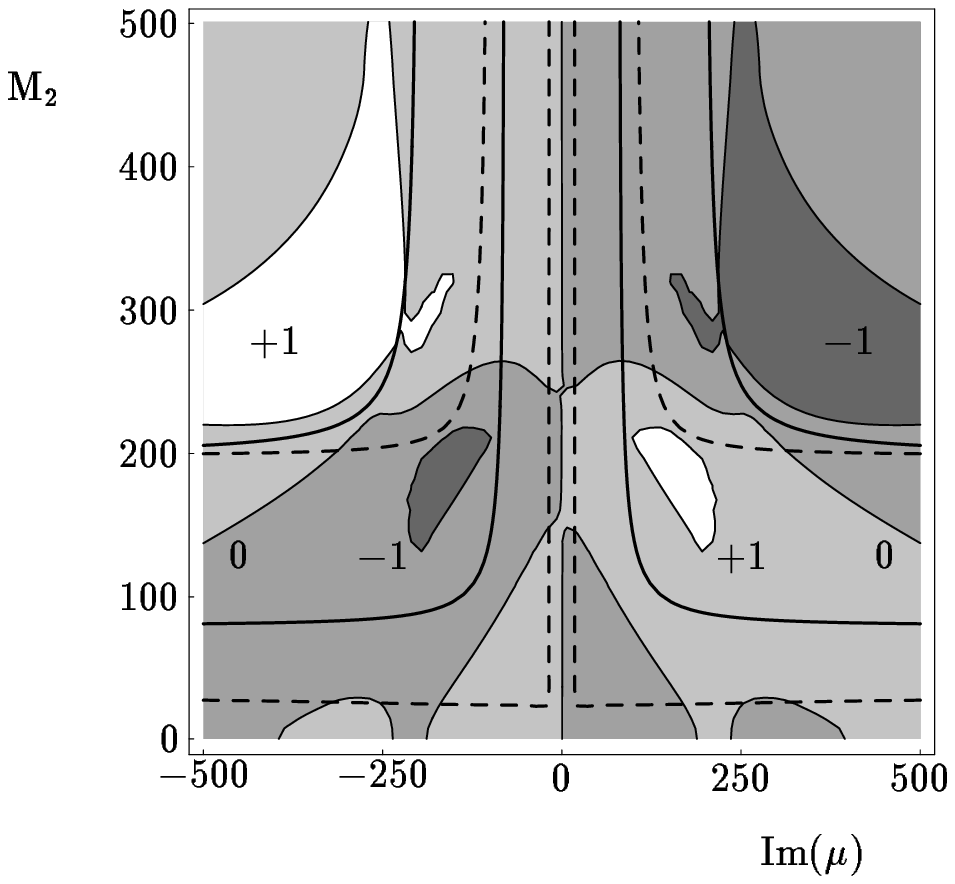,width=0.45\linewidth}    &    
\epsfig{file=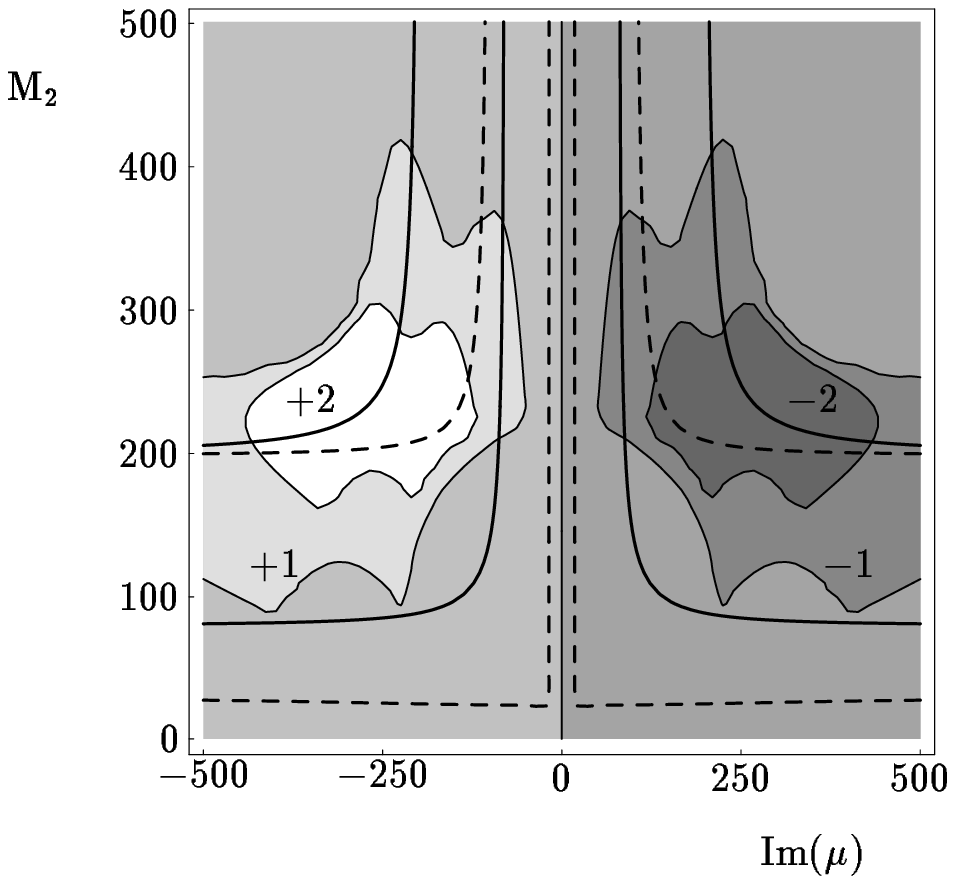,width=0.45\linewidth}    
\end{tabular}
\end{center}
\caption{\em Chargino contribution to the real and imaginary parts of the $t$
         EDFF and WEFF [in $10^{-3}\mu_t$ units] in the plane $M_2-{\rm Im}(\mu)$ 
         for $\tan\beta=1.6$, $|m^b_{LR}|=200$ GeV and $\varphi_{\tilde b}=\pi/2$
         at $\sqrt{s}=500$ GeV. 
         The lower (upper) solid isolines correspond to $m_{\tilde{\chi}^\pm_1}
         =90\ (200)$ GeV and the dashed isolines to $m_{\tilde{\chi}^0_1}=25\ 
         (100)$ GeV. \label{fig:4.2}}
\vspace*{0.3cm}
\end{figure}

As in the case of the neutralinos, the influence of chargino diagrams is 
enhanced for low $\tan\beta$.  

The results depend very little on $\varphi_{\tilde b}$ and mainly
on $\varphi_\mu$, the maxima being close to $\varphi_\mu=\pm\pi/2$.
For increasing $\tan\beta$ the dependence on $\varphi_{\tilde b}$ grows,
as it comes with a factor proportional to $m_b\tan\beta$.

The chargino contributions are the most important ones.
In Fig.~\ref{fig:4.2} the dependence on Im$(\mu)$ and $M_2$ is displayed
for $\tan\beta=1.6$. The only relevant CP--violating phase here has been set
to the most favorable case, $\varphi_\mu=\pi/2$ (the negative values of 
Im$(\mu)$ correspond to $\varphi_\mu=-\pi/2$). The symmetry with respect to
Im$(\mu)=0$ in Fig.~\ref{fig:4.2} reflects the (almost) independence of
$\varphi_{\tilde b}$, here set to $\pi/2$.\footnote{
All the contributions flip sign when the set ($\varphi_\mu,\varphi_{\tilde t},
\varphi_{\tilde b}$) is rotated by $\pi$. They vanish accordingly when all 
the phases are zero. See Fig.~\ref{fig:4.1} for illustration.} 
The same does not happen for the neutralino contributions for
a fixed value of $\varphi_{\tilde t}\ne 0,\pi$ in the plane $M_2-{\rm Im}(\mu)$.
The plots of Fig.~\ref{fig:4.2} exhibit a tendency to decoupling of the 
supersymmetric effects for increasing values of the mass parameters 
\cite{decoupling}.
The isolines for a couple of masses of the lightest charginos and neutralinos
in the same plane are also given for orientation. The current LEP2
experimental lower limits are $m_{\tilde{\chi}^0_1}\gsim 25$ GeV and 
$m_{\tilde{\chi}^\pm_1}\gsim 90$ GeV \cite{databook}.

The chargino contributions for the values $M_2=|\mu|=200$ GeV and
$\varphi_\mu=\pi/2$ are

\beq
d^\gamma_t[{\tilde \chi}^\pm]   &=& (0.869-1.870\ {\rm i})\times 10^{-3}\ \mu_t
\ ,\\
d^Z_t[{\tilde \chi}^\pm]        &=& (0.793-2.524\ {\rm i})\times 10^{-3}\ \mu_t
\ .
\eeq

\subsubsection{Gluinos and ${\tilde t}$ scalar quarks}

Their effect is roughly
proportional to $|m^t_{LR}|\sin\varphi_{\tilde t}$ times a chirality flipping
fermion mass, either $m_t$ or $M_3$ (the gluino mass). It is damped for
heavy gluinos circulating in the loop and also for large scalar quark masses 
due to decoupling. Both terms have opposite sign to Im($m^t_{LR}$) and the one
proportional to the gluino mass dominates.

The result for $M_2=200$ GeV and $\varphi_{\tilde t}=-\pi/2$ is

\beq
d^\gamma_t[{\tilde g}]  &=& (0.457+0.170\ {\rm i})\times 10^{-3}\ \mu_t \ , \\
d^Z_t[{\tilde g}]       &=& (0.155+0.059\ {\rm i})\times 10^{-3}\ \mu_t \ .
\eeq

\subsubsection{Total contribution}

\begin{figure}
\begin{center}
\epsfig{file=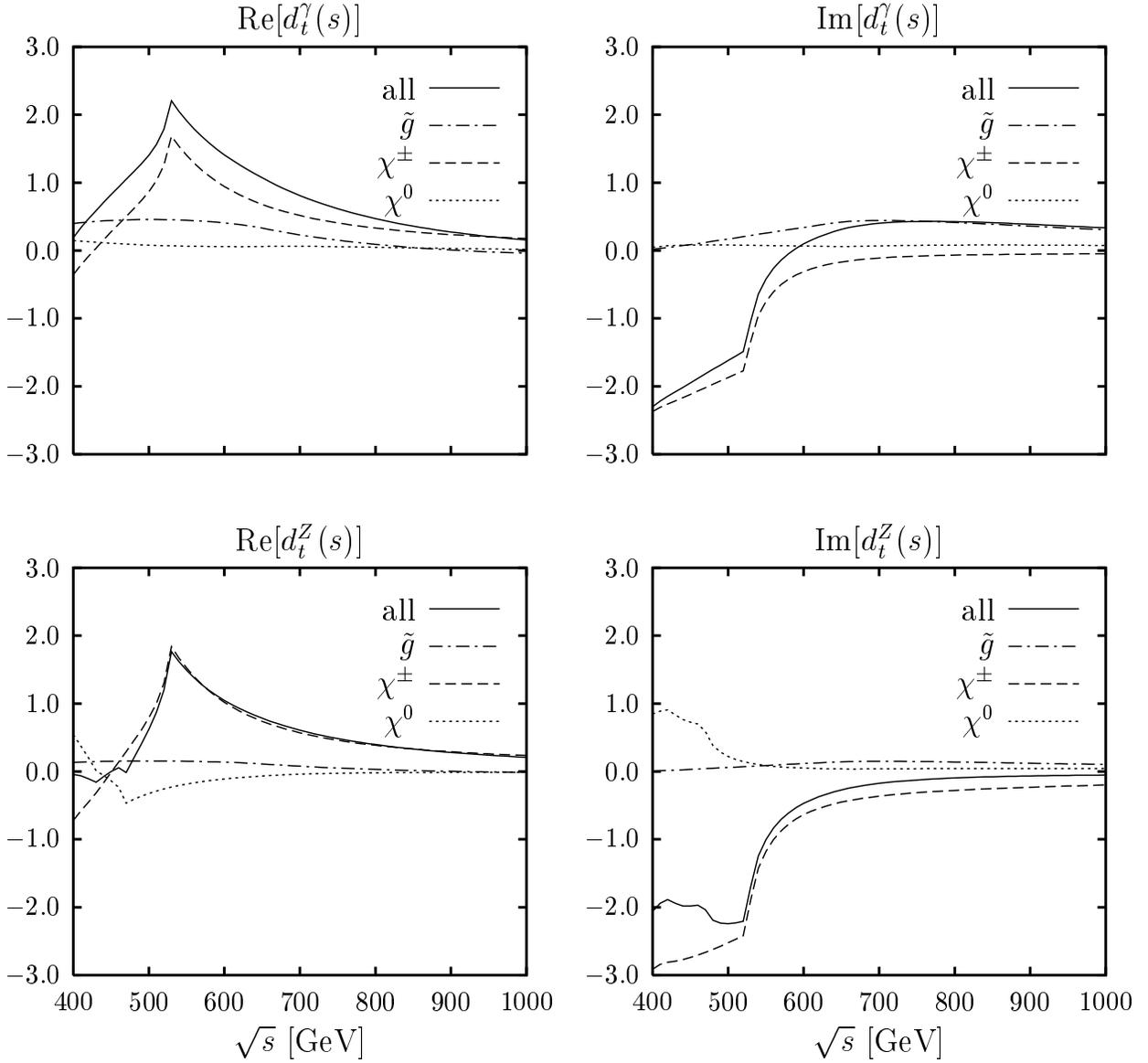,width=\linewidth}
\end{center}
\caption{\em The different contributions to the $t$ (W)EDFF [in $10^{-3}\mu_t$
units] for 
the reference set of SUSY parameters of Eq. (\ref{refset1}).\label{fig:4.3}}
\vspace*{0.3cm}
\end{figure}

In view of these results, we establish a set of SUSY parameters
and phases for which nearly all the contributions sum up constructively
at $\sqrt{s}=500$ GeV.
Our choice is
\beq
\mbox{Reference Set }\#1:& &\tan\beta=1.6 \nn \\
& &M_2=|\mu|=m_{\tilde{q}}=|m^t_{LR}|=|m^b_{LR}|=200 \mbox{ GeV} \nn \\
& &\varphi_\mu=-\varphi_{\tilde t}=-\varphi_{\tilde b}=\pi/2 ,
\label{refset1}
\eeq
for which the $t$ EDFF and WEFF reach the values:
\beq
d^\gamma_t      &=&     (1.407-1.618\ {\rm i})\times 10^{-3}\ \mu_t \\
d^Z_t           &=&     (0.624-2.242\ {\rm i})\times 10^{-3}\ \mu_t , 
\label{topff}
\eeq
at $\sqrt{s}=500$ GeV. 
To illustrate how the maximum effects appear we display in Fig.~\ref{fig:4.3}
the individual contributions as a function of $\sqrt{s}$.
The masses of the supersymmetric partners in the
loops are such that there are threshold enhancements in the vicinity of
$\sqrt{s}=500$ GeV. 

\section{CP--odd observables}
  
One needs to go beyond the $Z$ peak to produce $t$ quark pairs in $e^+e^-$
collisions and this implies to account for several features.
The photon--exchange diagram is no longer suppressed in this regime and 
therefore one needs to separate the contributions of electromagnetic and 
weak form factors \cite{bernre91b}.
Moreover, not only vertex-- but also box--diagrams correct the tree 
level process to one loop. The former corrections are parameterized by the 
electromagnetic and the weak vertex form factors but the latter demand the 
introduction of new form factors according to the more general topology of the 
process. 
For a realistic theory, one expects that any CP--odd observable 
will depend not only on the CP--violating effects due to
vertex corrections included in the electric and weak--electric dipole form
factors but also on possible CP--violating box contributions. This is indeed
the case of the supersymmetric models.

We concentrate on supersymmetric CP violating effects in $t$--pair
production at $e^+e^-$ colliders. 
CP conserving MSSM one--loop contributions to $e^+e^-\to f\bar f$ are discussed
in \cite{hs}.
To our knowledge, only the electric and the 
weak--electric form factors have been considered so far to parameterize these 
effects.\footnote{
A similar analysis for hadron colliders has been recently presented in 
Ref.~\cite{zhouhad}.}
Our purpose is
to evaluate the expectation value of several observables in the 
context of the MSSM with complex parameters to one loop. Eventually we
compare the contribution from the electric and weak--electric form
factors to these observables with the CP--violation effects coming from
the box contributions.

Our starting point is an initial CP--even eigenstate in the c.m.s. (laboratory
frame).
This is the case for unpolarized $e^+$ and $e^-$ beams but only
a good approximation in the case of longitudinally polarized beams 
\cite{bernre96a} (unless they are ideally 100$\%$ polarized): neglecting 
the electron mass, chirality conservation allows only 
interactions of equal helicity states that project into each other under a CP 
transformation.

Consider the process $e^+({\bf p_+})+e^-({\bf p_-})\to t({\bf k_+},\sone)+
\bar{t} ({\bf k_-},\stwo)$ (pair--production of polarized $t$ quarks). 
The decay channels labeled by $a$ and $c$ act as spin analyzers 
in $t + \bar t\to a(\qplus) + \bar c(\qminus)+X$.
The momenta and polarization vectors in the overall c.m.s. transform under
CP and CPT as follows:\footnote{T means reflection of spins and momenta.}
\beq
 \ba{rl}
 \mbox{CP}:& \ppmn\to -\pmpn=\ppmn \\
           & \kpmn\to -\kmpn=\kpmn \\
           & \qpmn\to -\qmpn \\
	   & \sone\leftrightarrow \stwo \\
 \ea
&\hspace{1cm}&
 \ba{rl}     
\mbox{CPT}:& \ppmn\to \pmpn=-\ppmn \\
           & \kpmn\to \kmpn=-\kpmn \\ 
           & \qpmn\to \qmpn \\
	   & \sone\leftrightarrow -\stwo 
 \ea
\eeq

From the unit momentum of one of the $t$ quarks in the c.m.s.
(say $\kplus$) and their polarizations 
($\sone$, $\stwo$) a basis of linearly independent CP--odd {\em spin
observables} can be constructed \cite{bernre91b}. They are 
classified according to their CPT parity. The spin observables are 
related to more realistic (directly measurable) {\em momentum observables}
based on the momenta of the top decay products 
\cite{bernre89b}. 
The polarizations can be analyzed through the angular correlation of the weak 
decay products, both in the nonleptonic and in the semileptonic channels:
\beq 
&&t(\kplusn)\to b({\bf q_b}) X_{\rm had}({\bf q_{X}})\ , \\  
&&t(\kplusn)\to b({\bf q_b})\ell^+(\qplus)\nu_\ell\quad (\ell=e,\mu,\tau)
\eeq
and the charged conjugated ones.\footnote{The leading QCD corrections to
$e^+e^-\to t\bar{t}$, that include a gluon emission, have a very small
effect on the $t$ spin orientation \cite{kodaira98}.} 
As $m_t>M_W+m_b$, the $t$ quark decays proceed
predominantly through $Wb$. Within the SM the angular distribution of the
charged lepton is a much better spin analyzer of the $t$ quark than that
of the $b$ quark or the $W$ boson arising from semileptonic or nonleptonic 
$t$ decays \cite{analtopspin}.
The dimensionless observables are easier to measure, for instance
the scalar CP--odd observables \cite{bernre91b}:
\beq
\hat{A}_1&\equiv&\pplus\cdot\dfrac{\qplusn\times\qminusn}
                                  {|\qplusn\times\qminusn|} 
 \hspace{1cm} \mbox{[CPT--even]} 
\label{A1}\\
\hat{A}_2&\equiv&\pplus\cdot(\qplusn+\qminusn)
 \hspace{1cm} \mbox{[CPT--odd]}
\label{A2}
\eeq 
or the CP--odd traceless tensors \cite{bernre91b}:
\beq
\hat{T}_{ij}&\equiv&(\qplusn-\qminusn)_i
       \frac{(\qplusn\times\qminusn)_j}{|\qplusn\times\qminusn|} + 
       (i\leftrightarrow j) \hspace{1cm} \mbox{[CPT--even]} 
\label{T33}\\
\hat{Q}_{ij}&\equiv&(\qplusn+\qminusn)_i(\qplusn-\qminusn)_j +
       (i\leftrightarrow j) \hspace{1cm} \mbox{[CPT--odd]}
\label{Q33}
\eeq
The reconstruction of the $t$ frame is not necessary
for the momentum observables. The observables
$\hat{A}_2$ and $\hat{Q}_{ij}$ do not involve angular correlations as
they could be measured considering separate samples of events in the
reactions $e^+e^-\to aX$ and $e^+e^-\to \bar{a}X$. Nevertheless it is 
convenient to treat them in an event--by--event basis \cite{bernre91b}.

One may obtain additional CP--odd observables from combinations
of the standard tensors by multiplying them with CP--even scalar weight 
functions to maximize the sensitivity to CP--violating effects ({\em optimal
observables}). Neglecting quartic terms in the CP--violating 
form factors, $\lambda_i$, the differential cross section can be written as
$\dd\sigma=\dd\sigma_0+\sum_i\lambda_i\dd\sigma^i_1$. It has been shown that 
the observables given by ${\cal O}_i=\dd\sigma^i_1/\dd\sigma_0$ have maximal 
sensitivity to the CP--violating terms $\lambda_i$ \cite{optimal}. The CP--odd 
and CPT--even (odd) observables are sensitive to the dispersive (absorptive)
parts of the CP--violating form factors. These include the real (imaginary) 
parts of the electric and weak--electric dipole form factors as well as
CP--violating contributions from box topologies.

Since the $t$ quark is a heavy fermion, the CP conjugate modes $t_L\bar t_L$ and
$t_R\bar t_R$ are produced with a sizeable rate. This allows to construct
the following CP--odd asymmetry \cite{peskin,ber95p}
\beq
A=\frac{\#(t_L\bar t_L)-\#(t_R\bar t_R)}{\#(t_L\bar t_L)+\#(t_R\bar t_R)}
\hspace{1cm} \mbox{[CPT--odd]}
\label{LR}
\eeq
This asymmetry is related to the one that can be
measured through the energy spectra of prompt leptons in the decays 
$t\to W^+b\to\ell^+\nu_\ell b$ and its conjugate.
The $W^+$ is predominantly longitudinally polarized and, assuming the standard 
$V$--$A$ interaction, the $b$ quark is preferentially
left--handed. The $W^+$ is mostly collinear with the $t$ polarization and
so is the $\ell^+$ anti--lepton. Above the $t\bar t$ threshold a
$\ell^+$ coming from a $t_R$ has more energy, due to the Lorentz boost, than
one produced in a $t_L$ decay. The same happens for the conjugate channel,
the  $\ell^-$ from a $\bar{t}_L$ is in average more energetic than the one
from the $\bar{t}_R$. Therefore, in the decay of the $t_R\bar{t}_R$ the 
anti--lepton from $t_R$ has a higher energy $E_+$, while in the decay of the 
pair $t_L\bar t_L$ the lepton from $\bar t_L$ has a higher energy $E_-$. 
Thus the asymmetry $A$ is sensitive to the energy asymmetry of the leptons
\cite{bernre89a} or $b$ quarks \cite{garfieldval} in the final state.

We will ignore possible CP violation in the $t$ or $\bar t$ decays.
CP--violating $t$ decays in supersymmetry have been considered in
\cite{aoki-oshimo}. We evaluate the influence on the CP--odd 
observables of the vertex and box diagrams through (ideal) spin observables.
The expectation value of the realistic momentum observables
given above will also be presented (assuming SM top decays) for comparison 
with experimental capabilities.

\section{MSSM full contribution to CP--odd observables}

\begin{figure}
\begin{center}
\epsfig{file=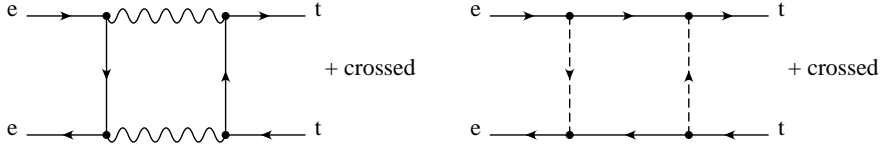,width=12cm}
\end{center}
\caption{\em Relevant generic box graphs for the MSSM at one loop.
\label{boxgraphs}}
\vspace*{0.3cm}
\end{figure}

The one--loop differential cross section for polarized $t$--pair production
in the MSSM \cite{hs} involves the box diagrams indicated in 
Fig.~\ref{boxgraphs} besides the vertex graphs of Fig.~\ref{fig1}. The class 
with vector boson exchange contains only SM contributions ($[eZZt]$, $[\nu_e 
WWb]$) whereas the other one is purely supersymmetric ($[\tilde e\tilde\chi^0
\tilde\chi^0\tilde t]$, $[\tilde\nu_e\tilde\chi^\pm\tilde\chi^\pm\tilde b]$) and
provides CP violating effects.
The box diagrams with Higgs boson exchange are proportional to the
electron mass and are neglected in the whole calculation (they are CP even 
in any case). When we refer to one--loop calculation in the following, the 
QED as well as the standard QCD corrections to the tree level process are 
excluded: they need real photonic and gluonic corrections to render an 
infrared finite result and constitute an unnecessary complication as they are 
CP--even and do not affect qualitatively our conclusions.

\subsection{Spin Observables}

\begin{table}
\caption{\em CP--odd spin observables and the coefficients for the expectation 
value of their integrated version
at $\sqrt{s}=500$ GeV, where only the CP--violating dipole form factors
are taken into account.
\label{tab6}}
\vspace{0.3cm}
\begin{center}
\begin{tabular}{|c|l|c|c|c|r|r|r|r|}
\hline
$i$ &CPT & ${\cal O}_i$ & {\bf a}&{\bf b} & $c_1$ & $c_2$ & $c_3$ 
                                     & $c_4$ \\
\hline \hline
1 & even& $(\soner-\stwor)_y$     & N$\uparrow$&N$\downarrow$
	& 0.526	& 0	& 1.517	& 0 	\\
2 & even& $(\soner\times\stwor)_x$& N$\uparrow$&L$\uparrow$	
	&$-0.465$& 0 	&$-0.061$& 0	\\
3 & even& $(\soner\times\stwor)_z$& N$\uparrow$&T$\uparrow$	
	& 0.708	& 0	& 0.144	& 0	\\
4 & odd & $(\soner-\stwor)_x$     & T$\uparrow$&T$\downarrow$
	& 0	& 0.930	& 0	& 0.123	\\
5 & odd & $(\soner-\stwor)_z$     & L$\uparrow$&L$\downarrow$
	& 0	&$-1.417$& 0	&$-0.287$\\
6 & odd & $(\soner\times\stwor)_y$& L$\uparrow$&T$\downarrow$
	& 0	& 0.263	& 0	& 0.758	\\
\hline
\end{tabular}
\end{center}
\vspace*{0.3cm}
\end{table}
	   
A list of CP--odd spin observables  
classified according to their CPT properties is shown in Table~\ref{tab6}.
Their expectation values as a function of $s$ and the scattering angle of the
$t$ quark in the overall c.m. frame are given by
\beq
\langle{\cal O}\rangle_{\mathbf{ab}} &=& \frac{1}{2\dd\sigma}\left[
 \sum_{\soner,\stwor=\pm \mathbf{a},\pm \mathbf{b}}
+\sum_{\soner,\stwor=\pm \mathbf{b},\pm \mathbf{a}}\right]
 \dd\sigma(\soner,\stwor)\ {\cal O}\ ,
\label{obsaverage} \\
\dd\sigma&=&\sum_{\pm\soner,\pm\stwor}\dd\sigma(\soner,\stwor)\ .
\label{spinaverage}
\eeq
The directions of polarization of $t$ and $\bar t$ ({\bf a}, {\bf b}) are 
taken normal to the scattering plane (N), transversal (T) or 
longitudinal (L). They can be either parallel ($\uparrow$) or antiparallel 
($\downarrow$) to the axes defined by $\hat{z}={\bf k_+}$, $\hat{y}={\bf k_+}
\times {\bf p_+}/|{\bf k_+}\times{\bf p_+}|$ and $\hat{x}=\hat{y}\times\hat{z}$.
Notice that
\beq
\langle{\cal O}^2\rangle_{\mathbf{ab}} &=& \frac{1}{\dd\sigma}
\sum_{\soner,\stwor=\pm \mathbf{a},\pm \mathbf{b}}
\dd\sigma(\soner,\stwor)\ {\cal O}^2
\eeq
and the same for the average of any CP--even quantity. If the information
of the $t$ scattering angle is not available one can consider integrated
observables 
\beq
\langle{\cal O}\rangle_{\mathbf{ab}} &=& \frac{1}{2\sigma}\left[
 \sum_{\soner,\stwor=\pm \mathbf{a},\pm \mathbf{b}}
+\sum_{\soner,\stwor=\pm \mathbf{b},\pm \mathbf{a}}\right]
 \sigma(\soner,\stwor)\ {\cal O}\ .
\eeq

\begin{figure}
\begin{center}
\epsfig{file=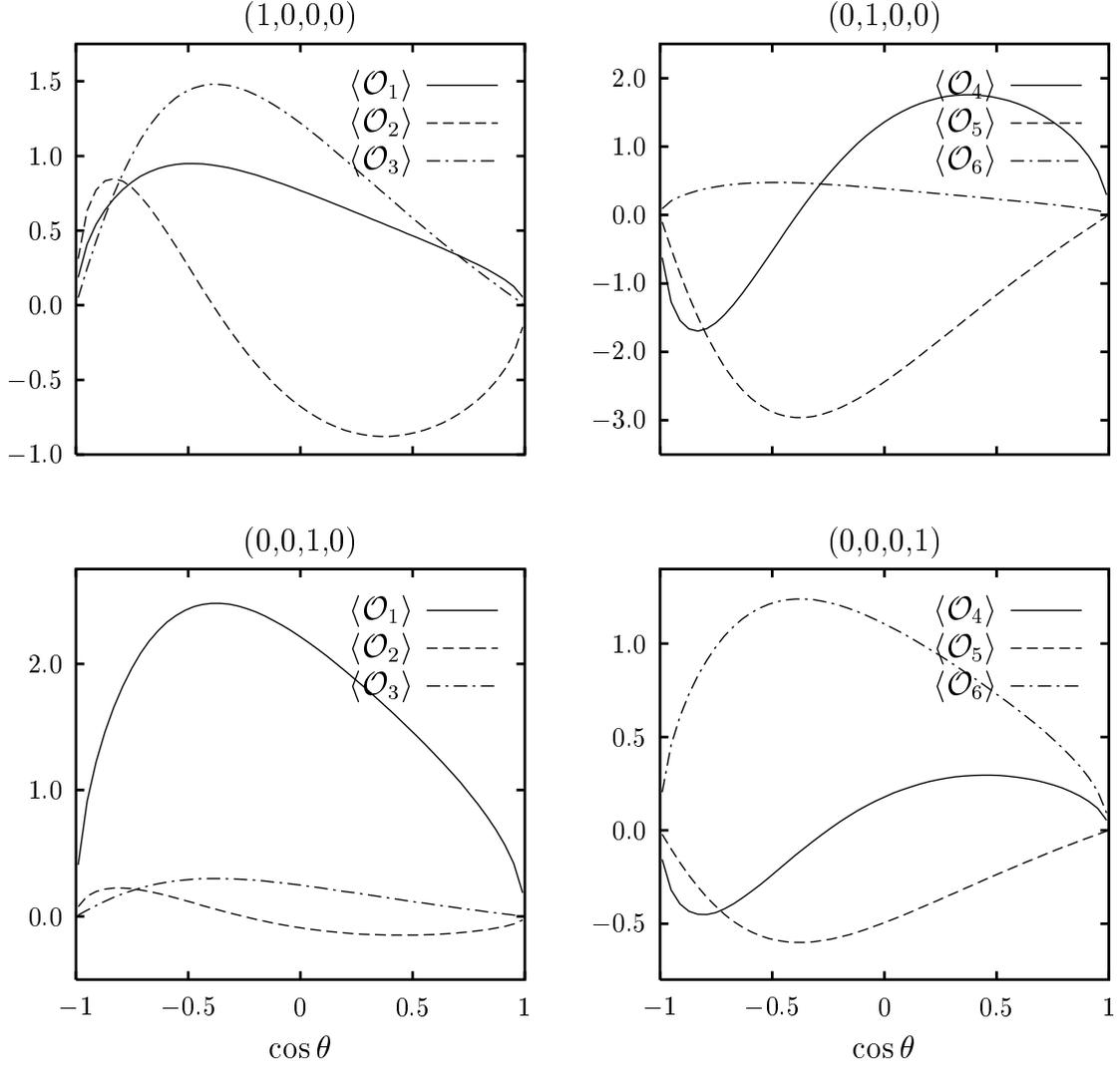,width=0.9\linewidth}
\end{center}
\caption{\em Dipole contributions to the expectation values of the spin 
observables at $\sqrt{s}=500$ GeV for different values of $({\rm
Re}[d^\gamma_t], 
\ {\rm Im}[d^\gamma_t],\ {\rm Re}[d^Z_t],\ {\rm Im}[d^Z_t])$ in $\mu_t$ units. 
\label{fig6.1}}
\vspace*{0.3cm}
\end{figure}

The contributions to the CP--odd observables are linear in the $t$ EDFF and 
WEDFF and in the CP--violating parts of the one--loop box graphs. 
The shape of the different dipole contributions to these observables 
is depicted in Fig.~\ref{fig6.1}. 
The coefficients of the dipoles in the linear expansion of the
integrated spin observables,
\beq
\langle{\cal O}\rangle_{\mathbf{ab}}\equiv\frac{2m_t}{e}\left(
  c_1\ {\rm Re}[d^\gamma_t]
+ c_2\ {\rm Im}[d^\gamma_t]
+ c_3\ {\rm Re}[d^Z_t]
+ c_4\ {\rm Im}[d^Z_t]\right)\ ,
\eeq
are shown in 
Table~\ref{tab6} for their integrated version at $\sqrt{s}=500$ GeV.
They are model independent.

\begin{figure}
\begin{center}
Reference Set $\#1$ \\ ~ \\
\epsfig{file=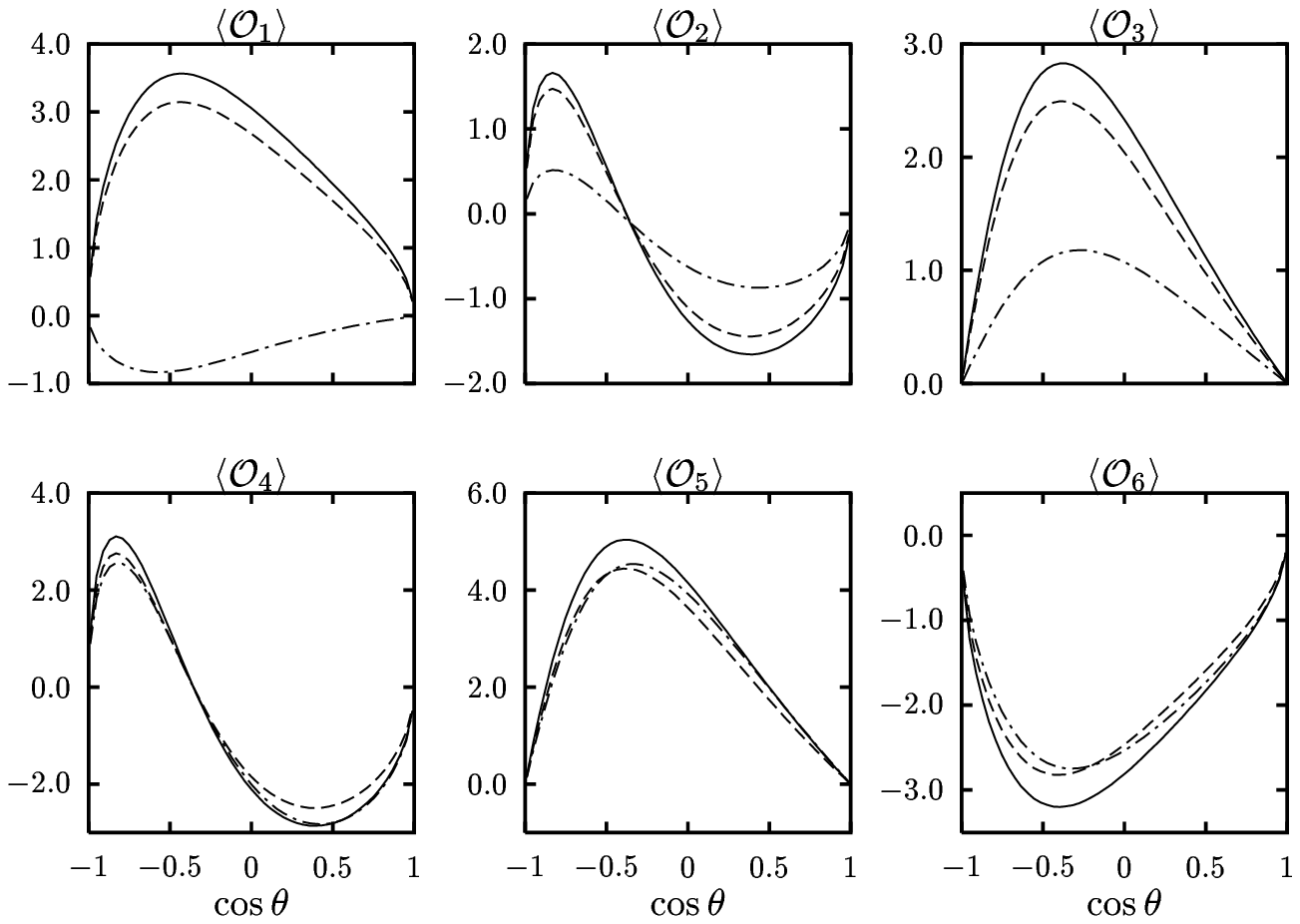,width=0.65\linewidth}
\\ ~ \\
Reference Set $\#2$ \\ ~ \\
\epsfig{file=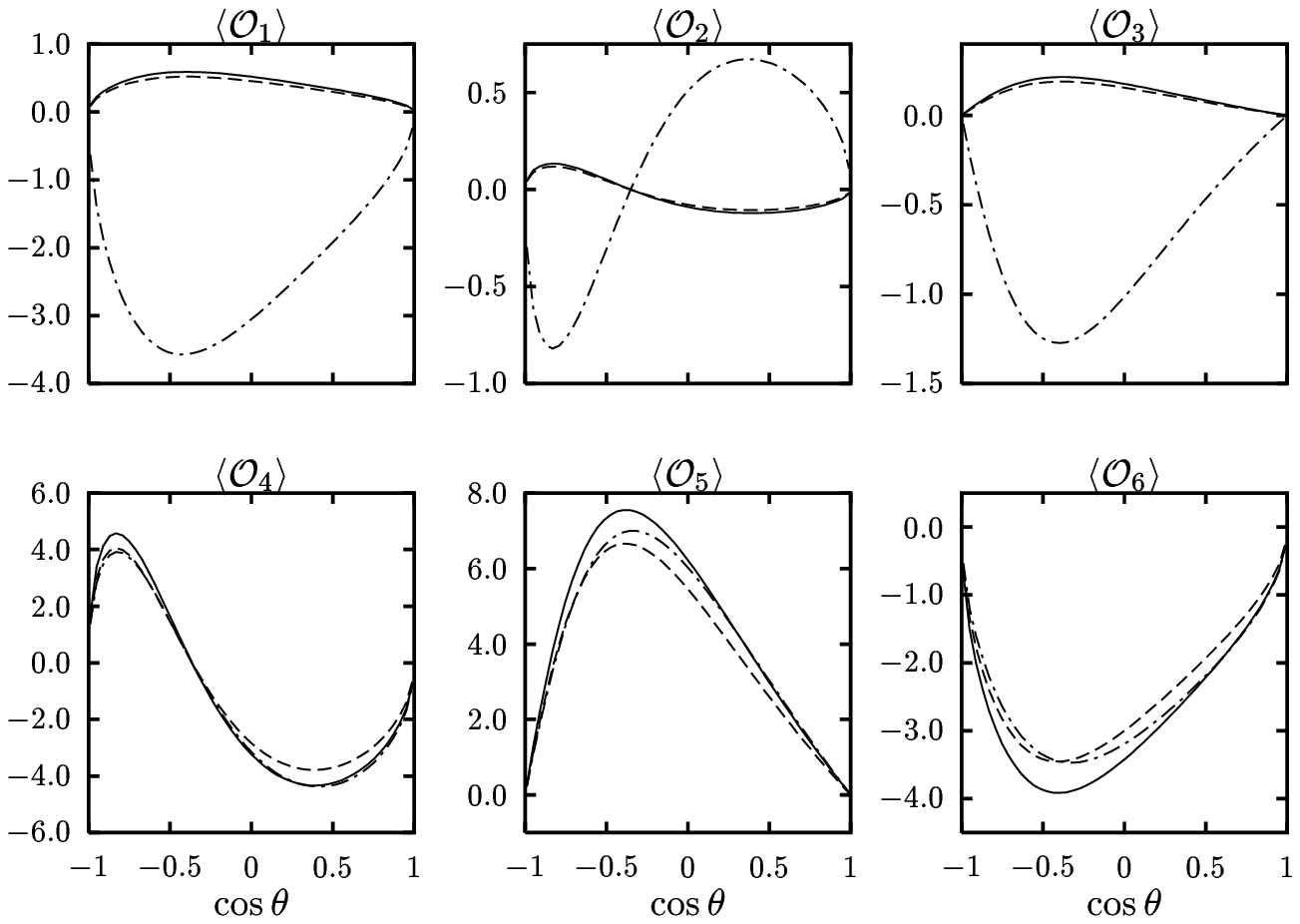,width=0.65\linewidth}
\end{center}
\caption{\em Expectation value of the spin observables [in $10^{-3}$ units] for 
the two reference sets of SUSY parameters, assuming for the cross section: 
  (i) tree level plus contribution from (W)EDFFs only (solid line);  
 (ii) one loop including all the vertex corrections and the self--energies
      (dashed line);
(iii) complete one loop (dot--dashed line).
\label{fig2}}
\vspace*{0.3cm}
\end{figure}

In Fig.~\ref{fig2} we compare the contributions of dipoles and boxes to
the spin observables, for two reference set of parameters. The first set
was given in (\ref{refset1}) and is the one that has been shown to enhance 
the dipole effects. The second one is:
\beq
\mbox{Reference Set }\#2:& &\tan\beta=1.6 \nn \\
& &M_2=|\mu|=m_{\tilde{q}}=|m^t_{LR}|=|m^b_{LR}|=200 \mbox{ GeV} \nn \\
& &\varphi_\mu=\varphi_{\tilde t}=\varphi_{\tilde b}=\pi/2\ ,
\label{refset2}
\eeq
where, compared to (\ref{refset1}), only the CP--violating phases have been 
changed.
The shape of the solid and dashed curves is the same in all cases, as expected. 
Their different size is due to the contributions to the normalization factors 
coming from self--energies, A(W)MFFs and other CP--even corrections.
The plots show that for the Set $\#1$ the MSSM box graphs contribute 
(in the case of CPT--even observables) to CP violation in the process 
$e^+e^-\to t \bar t$ by roughly the 
same amount and with a different profile than the EDFFs and WEDFF of the MSSM. 
For nearly all cases this eventually results in a large reduction of the value 
of any CP--odd observable with respect to the expectations from the dipoles
alone, as it will be shown below. Other sets of SUSY 
parameters, that do not enhance the dipole contributions, can provide
instead {\em smaller dipole form factors but larger observable effects}. This
is the case of the Set $\#2$, that yields much smaller values for the 
real parts of the dipole form factors at $\sqrt{s}=500$ GeV than we had before 
in Eq.~(\ref{topff}),
\beq
d^\gamma_t	&=&	(0.547-1.889\ {\rm i})\times 10^{-3}\ \mu_t , \\
d^Z_t		&=&	(0.362-2.319\ {\rm i})\times 10^{-3}\ \mu_t . 
\label{topff2}
\eeq
As a consequence, the CPT--even observables receive their contributions
mainly from the box diagrams (Fig.~\ref{fig2}) for the Set $\#2$. 

\begin{table}
\caption{\em Ratio $r\equiv\langle{\cal O}\rangle/\sqrt{\langle{\cal O}^2\rangle}$
 [in $10^{-3}$ units] of the integrated spin observables at $\sqrt{s}=500$ GeV
 for the two reference sets of SUSY parameters. The left column excludes the box 
 corrections and the right one comes from the complete one--loop cross section 
 for $e^+e^-\to t\bar t$.\label{tab7}}
\vspace{0.3cm}
\begin{center}
\begin{tabular}{|c|l|c|c|c|r|r|r|r|}
\hline
$i$ &CPT & ${\cal O}_i$ & {\bf a}&{\bf b} & 
\multicolumn{2}{|c|}{Set $\#1$} &
\multicolumn{2}{|c|}{Set $\#2$} \\
\hline \hline
1 & even&  $(\soner-\stwor)_y$ & N$\uparrow$&N$\downarrow$
	& 1.216 &$-0.231$& 0.207 &$-1.394$      \\
2 & even&  $(\soner\times\stwor)_x$& N$\uparrow$&L$\uparrow$	
	&$-0.755$&$-0.489$&$-0.053$&$0.318$       \\
3 & even&  $(\soner\times\stwor)_z$& N$\uparrow$&T$\uparrow$	
	& 1.184	&$0.625$&0.090	&$-0.598$       \\
4 & odd &  $(\soner-\stwor)_x$     & T$\uparrow$&T$\downarrow$
	&$-1.230$&$-1.421$&$-1.888$&$-2.217$ \\
5 & odd &  $(\soner-\stwor)_z$     & L$\uparrow$&L$\downarrow$
	& 2.550 &$2.739$ & 3.823 &4.216   \\
6 & odd &  $(\soner\times\stwor)_y$& L$\uparrow$&T$\downarrow$
	&$-1.683$&$-1.751$&$-2.050$&$-2.216$ \\
\hline
\end{tabular}
\end{center}
\vspace*{0.3cm}
\end{table}

The previous arguments are reflected in Table~\ref{tab7} where the
ratio $r\equiv\langle{\cal O}\rangle/\sqrt{\langle{\cal O}^2 
\rangle}$ is shown for the integrated spin observables.
We compare the result when only the self energies and vertex corrections
are included (left column) with the complete one--loop calculation (right 
column). The shape of the observables as a function of the $t$
polar angle (Fig.~\ref{fig2}) is such that signal for CP--violation $r$ is 
still not very large for a couple of CPT--even observables with the Set $\#2$ 
but it is much better for all the others. This illustrates that the dipole form
factors of the $t$ quark are not sufficient to parameterize observable 
CP--violating effects and the predictions can be wrong by far.

As a final remark, note that $\langle{\cal O}_5\rangle=A$, the 
asymmetry defined in Eq.~(\ref{LR}).
Attending to Table~\ref{tab6} it happens to be the most sensitive observable
to the imaginary part of the EDFF and also the best one to test CP violation
for our choice of SUSY parameters (Table~\ref{tab7}).
The observables $(\soner-\stwor)$ can still give information on CP violation 
when the polarization of one of the $t$ quarks is not analyzed.\footnote{
Anyway a comparison between two samples, one with polarized
$t$ and the other with polarized $\bar t$, is necessary to build the
genuine CP--odd observable of Eq.~(\ref{spinaverage}).} The sensitivity
of the single--spin polarization to CP--violation is indeed worse: 
$\langle{\cal(\soner-\stwor)}_{x,y,z}\rangle_{\bf ab}
=2\langle{\cal(\soner)}_{x,y,z}\rangle_{\bf a}$ for {\bf a}({\bf b})=
N$\uparrow$($\downarrow$),T$\uparrow$($\downarrow$),L$\uparrow$($\downarrow$).

\subsection{Momentum Observables}

Consider now the decay channels labeled by $a$ and $c$ acting as spin analyzers 
in $t + \bar t\to a(\qplus) + \bar c(\qminus)+X$. 
The expectation value of a CP--odd observable is given by the average over the 
phase space of the final state particles,
\beq
\langle{\cal O}\rangle_{ac} = \frac{1}{2}\left[ 
\langle{\cal O}\rangle_{a\bar c} + \langle{\cal O}\rangle_{c\bar a} \right] =
\frac{1}{2\sigma_{ac}}\int
\left[ \dd\sigma_{a\bar c} + \dd\sigma_{c\bar a} \right]\ {\cal O}\ ,
\eeq
where both the process ($a\bar c$) and its CP conjugate ($c\bar a$) are 
included and 
\beq
\sigma_{ac}=\int\dd\sigma_{a\bar c}=\int\dd\sigma_{c\bar a} \ ,
\eeq
in full analogy with Eqs.~(\ref{obsaverage},\ref{spinaverage}).
The differential cross section for $t$--pair production and decay is
evaluated for every channel using the narrow width approximation.

\begin{table}
\caption{\em Ratio $r$ [in $10^{-3}$ units] of the momentum observables 
(\ref{A1}--\ref{Q33})  
at $\sqrt{s}=500$ GeV for three different channels: $t + \bar t\to a(\qplus) + 
\bar c(\qminus)+X$, given the Set $\#2$ of SUSY parameters 
(\ref{refset2}). The left column includes only the $t$ (W)EDFF corrections  
and the right one comes from the complete one--loop cross section for 
$e^+e^-\to t\bar t$. \label{tab9}}
\vspace{0.3cm}
\begin{center}
\begin{tabular}{|l|c|r|r|r|r|r|r|}
\hline
CPT	& ${\cal O}$ & \multicolumn{2}{|c|}{$b-b$} 
	& \multicolumn{2}{|c|}{$\ell-b$} 
	& \multicolumn{2}{|c|}{$\ell-\ell$} \\ 
\hline \hline
even& $\hat{A}_1$   &
$-0.036$	&0.242 		&
0.030		&$-0.202$ 	& 	
0.068	&$-0.467$\\
odd & $\hat{A}_2$ 	&  
0.270		&0.304 	&
$-0.180$	&$-0.204$& 	
$-0.501$	&$-0.812$\\
even& $\hat{T}_{33}$&
$-0.006$	&$0.042$&
0.021	&$-0.140$& 	
$-0.037$	&$0.248$\\
odd & $\hat{Q}_{33}$&
0.486		&0.542 	& 
$-0.335$	&$-0.374$& 	
$-1.274$	&$-1.420$\\
\hline
\end{tabular}
\end{center}
\vspace*{0.3cm}
\end{table} 

In Table~\ref{tab9} the ratio $r$ is shown for three different decay channels 
at $\sqrt{s}=500$ GeV and some {\em realistic} CP--odd observables involving 
the momenta of the decay products analyzing $t$ and $\bar t$ polarizations 
in the laboratory frame. The leptonic
decay channels are the best $t$ spin analyzers but the number of leptonic
events is also smaller. The Reference Set $\#2$ has 
been chosen. As expected, the dipole contributions (left columns) to the CPT 
even observables are very small for this choice of SUSY parameters 
but the actual expectation values (right columns) are larger.

The previous results were obtained for unpolarized electron and positron beams.
Let $P_\pm$ be the degree of longitudinal polarization of the initial
$e^\pm$. The differential cross section reads now
\beq
\dd\sigma=\frac{1}{4}\left[(1+{\rm P}_+)(1+{\rm P}_-)\ \dd\sigma_R\ +\ 
	     (1-{\rm P}_+)(1-{\rm P}_-)\ \dd\sigma_L\right]\ ,
\label{sigmapol}
\eeq
where $\sigma_{R/L}$ corresponds to the cross section for electrons and
positrons with equal right/left--handed helicity. Chirality conservation
suppresses opposite helicities. Table~\ref{tab10} summarizes some extreme
cases. If both beams are fully polarized, ${\rm P}_+={\rm P}_-=\pm 1$, the 
ratio $r$ is the same as for (${\rm P}_\pm=0$, ${\rm P}_\mp=\pm 1$), 
respectively 
(Table~\ref{tab10}), but the cross sections are twice as much (\ref{sigmapol}),
which results in a higher statistical significance of the CP signal. From 
comparison of Table~\ref{tab9} (right columns) with Table~\ref{tab10}
is clear that left--handed polarized beams enhance the sensitivity to 
CP--violating effects.

\begin{table}
\caption{\em The same as in Table~\ref{tab9} assuming the complete one--loop 
cross section for $e^+e^-\to t\bar t$ and longitudinal polarizations for
one of the $e^\pm$ beams, ${\rm P}_\pm$. \label{tab10}}
\vspace{0.3cm}
\begin{center}
\begin{tabular}{|c|r|r|r|}
\hline
\multicolumn{4}{|c|}{${\rm P}_\pm=0$, ${\rm P}_\mp=-1$; 
		$\sigma_{t\bar t}=0.707$ pb} \\
\hline
${\cal O}$ & $b-b$ & $\ell-b$ & $\ell-\ell$ \\
\hline \hline
$\hat{A}_1$   &
0.359       &
$-0.298$       &
$-0.667$        \\
$\hat{A}_2$       &
0.397       &
$-0.254$       &
$-0.952$        \\
$\hat{T}_{33}$&
0.065       &
$-0.214$       &
0.383        \\
$\hat{Q}_{33}$&
0.708       &
$-0.488$       &
$-1.848$        \\
\hline
\end{tabular}
\begin{tabular}{|c|r|r|r|}
\hline
\multicolumn{4}{|c|}{${\rm P}_\pm=0$, ${\rm P}_\mp=+1$; $\sigma_{t\bar t}=0.355$ pb} \\
\hline
${\cal O}$ & $b-b$ & $\ell-b$ & $\ell-\ell$ \\
\hline \hline
$\hat{A}_1$   &
$0.016$       &
$-0.009$       &
$-0.053$        \\
$\hat{A}_2$       &
0.134       &
$-0.099$       &
$-0.468$        \\
$\hat{T}_{33}$&
$-0.003$       &
$0.006$       &
$-0.020$        \\
$\hat{Q}_{33}$&
0.210       &
$-0.145$       &
$-0.555$        \\
\hline
\end{tabular}
\end{center}
\end{table}
Concerning the experimental reach of this analysis,
the statistical significance of the signal of CP violation is given
by $S=|r|\sqrt{N}$ with $N=\epsilon{\cal L}\sigma_{t\bar t}\mbox{BR}
(t\to a)\mbox{BR}(\bar t\to\bar c)$ where $\epsilon$ is the detection
efficiency and ${\cal L}$ the integrated luminosity of the collider.
The branching ratios of the $t$ decays are BR $\simeq 1$ for the $b$ channel
and BR $\simeq 0.22$ for the leptonic channels ($\ell=e,\mu$).
At $\sqrt{s}=500$ GeV the total cross section for $t$--pair production is
$\sigma_{t\bar t}\simeq 0.5$ pb and the NLC integrated luminosity ${\cal L}
\simeq 50$ fb$^{-1}$ \cite{nlc}. Assuming a perfect detection efficiency, one
gets $\sqrt{N}\simeq 160, 75, 35$ for the channels $b-b$, $\ell-b$,
$\ell-\ell$, respectively. With these statistics, values of $|r|\sim 10^{-2}$
would be necessary to achieve a 1 s.d. effect, which does not seem to
be at hand in the context of the MSSM, even for polarized beams, as 
Tables~\ref{tab9} and \ref{tab10} show. 


\section{Summary and conclusions}

The one--loop expressions of the dipole form factors of fermions in terms 
of arbitrary complex couplings in a general renormalizable theory for the 
't Hooft--Feynman gauge have been given. The CP--violating (--conserving)
dipole form factors depend explicitly on the imaginary (real) part of 
combinations of the couplings. 

As an application, the 
electric and the weak--electric dipole form factors of the $t$ quark
(for $s>4m^2_t$) have been evaluated for the MSSM with preserved R--parity
and non--universal soft--breaking terms. A version with complex couplings
of the MSSM has been implemented in order to get one--loop CP--violating 
effects. The results depend on three physical CP phases.
The supersymmetric parameter space has been scanned in search for the
largest contributions to the dipoles. 
The $t$ dipole form factors are larger in the low $\tan\beta$ 
scenario. The values depend strongly on the interplay between the energy at 
which they are evaluated and the set of supersymmetric parameters used as
inputs. The real and imaginary parts can reach values of similar size: 
at $\sqrt{s}=500$ GeV, the $t$ EDFF and WEDFF are of ${\cal O}(10^{-19})\ e$cm.

Away from the $Z$ peak both the electromagnetic and the weak dipole form
factors are equally relevant but not enough to parameterize all the physical
effects (in particular CP--violation). The case of $t$--pair production
in high energy $e^+e^-$ colliders has been considered to illustrate this fact. 
Taking several CP--odd spin--observables based on the $t$ and $\bar t$
polarization vectors, the different contributions from vertex and box
corrections have been evaluated. There is no one loop contribution from the
SM sector. It has been shown that, for the set of supersymmetric parameters 
that provides sizeable values of the $t$ CP--violating dipoles, the SUSY box 
contributions happen to contribute with opposite sign and in a similar 
magnitude, yielding altogether a much smaller CP--violating observable effect.
Another configuration has been shown for which the dipoles are
smaller but the combined effect is larger. The same analysis has been performed
using instead realistic observables based on the momenta of $t$ and $\bar t$ 
decay products, with similar results. The channels in which $bb$, $b\ell$ 
and $\ell\ell$ act as spin analyzers have been used under the assumption of 
standard CP--conserving decays and using the narrow width approximation. As 
expected, the leptons are the best $t$ spin analyzers yielding the maximal 
values for the signal of CP--violation, $r\simeq 0.5\times 10^{-3}$. 
Nevertheless the statistics for such events is smaller.


\section*{Acknowledgements}

We wish to thank T. Hahn for very valuable assistance in the optimization of 
the computer codes and the preparation of the figures. 
Useful discussions with J. Bernab\'eu and A. Masiero are also gratefully 
acknowledged. 
One of us (S.R.) would like to thank the INFN Theoretical Group of Padua for 
the nice and fruitful hospitality enjoyed during the preparation of part 
of this paper. S.R. has been supported by the EC under contract 
ERBFMBICT972474.
J.I.I. was supported by the Fundaci\'on Ram\'on Areces at the Karlsruhe U., 
where most of this work has been carried out, and partially by the 
Spanish CICYT under contract AEN96-1672.

\appendix


\section{Decomposition of the one-loop 3--point integrals 
\label{appendix-a}}

\begin{figure}[htb]
\begin{center}
\begin{tabular}{c}
\epsfig{figure=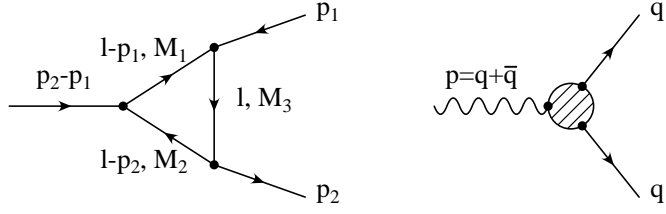,height=3cm}
\end{tabular}
\end{center}
\caption{\em Momentum convention for 3--point integrals.
\label{fig:3pf}}
\vspace*{0.3cm}
\end{figure}
The systematic way of dealing with one-loop integrals consists of
reducing the tensor integrals to scalar ones. We employ the following 
set of 3--point tensor integrals (see Fig.~\ref{fig:3pf}):
\beq
{C}_{\{0,\ \mu,\ \mu\nu\}}(p_1,p_2,M_1,M_2,M_3)\equiv\frac{16\pi^2}{i}
\mu^{4-D}\int\frac{\dd^D l}{(2\pi)^D}\frac{\{1,\ l_\mu,\ l_\mu l_\nu\}}
{{\cal D}_1{\cal D}_2{\cal D}_3}\ ,
\label{c}
\eeq
with
\beq
{\cal D}_1&=&(l-p_1)^2-M^2_1+i\epsilon\ , \nonumber\\
{\cal D}_2&=&(l-p_2)^2-M^2_2+i\epsilon\ , \nonumber\\
{\cal D}_3&=&l^2-M^2_3+i\epsilon\ .
\eeq
and the orthogonal reduction ($k_{\pm}=p_1\pm p_2$) \cite{topol},
\beq
{C}_\mu&\equiv&C^+_1 k_{+\mu}+C^-_1 k_{-\mu} \ , \\
{C}_{\mu\nu}&\equiv&C^+_2 k_{+\mu}k_{+\nu}+C^-_2 k_{-\mu}k_{-\nu} 
        +C^{+-}_2 [k_{+\mu}k_{-\nu}+k_{+\nu}k_{-\mu}]+C^0_2 g_{\mu\nu}\ .
\eeq

This decomposition is very convenient because the integral
(\ref{c}) is invariant under the combined replacements 
$p_1\leftrightarrow p_2$ and $M_1\leftrightarrow M_2$, and we are 
dealing with equal mass final-state particles $p_1^2=p_2^2=m^2$.
Therefore $C^-_1$ and $C^{+-}_2$ are antisymmetric under these
replacements while the other scalar integrals are symmetric, and the
case $M_1=M_2$ automatically yields $C^-_1=C^{+-}_2=0$.

For comparison we also list the tensor decomposition defined in
\cite{Denner93}. 
\beq
{C}^{(D)}_{\{0,\ \mu,\ \mu\nu\}}(p_1,p_2,M_3,M_1,M_2)\equiv\frac{16\pi^2}{i}
\mu^{4-D}\int\frac{\dd^D l}{(2\pi)^D}\frac{\{1,\ l_\mu,\ l_\mu l_\nu\}}
{{\cal D}_1{\cal D}_2{\cal D}_3}\ ,
\eeq
with
\beq
{\cal D}_1&=&l^2-M^2_3+i\epsilon\ , \nonumber\\
{\cal D}_2&=&(l+p_1)^2-M^2_1+i\epsilon\ , \nonumber\\
{\cal D}_3&=&(l+p_2)^2-M^2_2+i\epsilon\ .
\eeq
and the tensor decomposition
\beq
{C}^{(D)}_\mu&\equiv&{p_1}_\mu C^{(D)}_1 + {p_2}_\mu C^{(D)}_2\ ,\\
{C}^{(D)}_{\mu\nu}&\equiv&g_{\mu\nu} C^{(D)}_{00}
               + {p_1}_\mu{p_1}_\nu C^{(D)}_{11}
               + {p_2}_\mu{p_2}_\nu C^{(D)}_{22}
               + ({p_1}_\mu{p_2}_\nu + {p_1}_\nu{p_2}_\mu)
               C^{(D)}_{12}\ .\nonumber\\&&
\eeq
These scalar integrals are related to the ones obtained by orthogonal
reduction in the following way:
\begin{eqnarray}
C^{(D)}_1 & = & -C_1^+ -C_1^-,
\\
C^{(D)}_2 & = & -C_1^+ +C_1^-,
\\
C^{(D)}_{00} & = & C_2^0,
\\
C^{(D)}_{11} & = & C_2^+ +C_2^- +2C_2^{+-},
\\
C^{(D)}_{22} & = & C_2^+ +C_2^- -2C_2^{+-},
\\
C^{(D)}_{12} & = & C_2^+-C_2^-,
\end{eqnarray}
with the arguments
$$
C_i^{(D)} = C_i^{(D)}(p_1,p_2,M_3,M_1,M_2),\quad
C_i^j = C_i^j(-p_1,-p_2,M_1,M_2,M_3).
$$


\section{Conventions for fields and couplings in the SM and the MSSM
         \label{appendix-c}}

  \subsection{The conventions}

We use the conventions of Ref.~\cite{mssm}. The covariant 
derivative acting on a SU(2)$_L$ weak doublet field with hypercharge $Y$ is 
given by
\beq
D_\mu=\partial_\mu+{\rm i}g\frac{\tau^a}{2}W^a_\mu+{\rm i}g'\frac{Y}{2}B_\mu \ ,
\label{covder}
\eeq
where $\tau^a$ are the usual Pauli matrices and the electric charge
operator is $Q_f=I^f_3+Y/2$, with $I^f_3=\tau^3/2$. The $Z$ and
photon fields are defined by
\beq
Z_\mu&=&W^3_\mu\cos\theta_W-B_\mu\sin\theta_W \ ,\\
A_\mu&=&W^3_\mu\sin\theta_W+B_\mu\cos\theta_W \ .
\eeq
The charged weak boson field is $W^\pm_\mu=\frac{1}{\sqrt{2}}
(W^1_\mu\mp {\rm i} W^2_\mu)$.

In the Standard Model (SM), there is only one Higgs doublet ${\bf H}$ with 
hypercharge $Y=1$. After spontaneous symmetry breaking (SSB), the physical 
Higgs field $H^0$ and the would--be--Goldstone bosons $\chi$ and $\phi^\pm$ are
given by
\beq
{\bf H}=\left(\ba{c} \phi^+ \\ \frac{1}{\sqrt{2}}[v+H^0+{\rm i}\chi] 
\ea\right)\ .
\eeq

In the Minimal Supersymmetric Standard Model (MSSM) there are the same matter
and gauge field multiplets as in the SM, each supplemented
by superpartner fields to make up complete supersymmetry
multiplets. The matter and gauge fields have the same quantum number
assignments as in the SM. But in the MSSM there are two
Higgs doublets with opposite hypercharges. Each forms a chiral 
supersymmetry multiplet and an SU(2) doublet. 
The Lagrangian we use is defined in Ref.~\cite{Haber93}, especially in
Eq.~(1.45) and Eq.~(1.54). We do not impose any reality constraint
onto the parameters except for the reality of the
Lagrangian.

Spontaneous breakdown of the SU(2)$\times$U(1) gauge symmetry leads to the 
existence of five physical Higgs particles: two CP-even Higgs bosons $h$ and 
$H$, a CP-odd or pseudoscalar Higgs boson $A$, and two charged Higgs particles 
$H^\pm$. They are grouped in two doublets,
${\bf H}_1\equiv i\tau^2{\bf\Phi}_1^*$ and ${\bf H}_2\equiv{\bf\Phi}_2$,
with opposite  hypercharges ($Y=\mp 1$), where  
\beq
{\bf\Phi}_1=\left(\ba{c} \phi^+_1\\ \phi^0_1 \ea\right)\ , \ \
{\bf\Phi}_2=\left(\ba{c} \phi^+_2 \\ \phi^0_2 \ea\right)\ . 
\eeq
After SSB these are expressed in terms of the physical fields $h$, $H$, $A$, 
$H^\pm$ and the would-be-Goldstone bosons $G^0$, $G^\pm$ by
\beq
\phi^+_1&=&-H^+\sin\beta+G^+\cos\beta \ , \\
\phi^0_1&=&\frac{1}{\sqrt{2}}\left\{v_1+[(-h\sin\alpha+H\cos\alpha)+
                        {\rm i}(-A\sin\beta+G^0\cos\beta)]\right\} \ , \\
\phi^+_2&=& H^+\cos\beta+G^+\sin\beta \ , \\
\phi^0_2&=&\frac{1}{\sqrt{2}}\left\{v_2+[( h\cos\alpha+H\sin\alpha)+
                        {\rm i}( A\cos\beta+G^0\sin\beta)]\right\} \ . 
\eeq

Besides the four masses, two additional parameters are needed to describe the 
Higgs sector at tree-level: $\tan\beta=v_2/v_1$, the ratio of the two 
vacuum expectation values, and a mixing angle $\alpha$ in the CP-even sector. 
However, only two of these parameters are independent. Using $M_A$ and 
$\tan\beta$ as input parameters, the masses and the mixing angle $\alpha$ in 
the $H,h$ sector read \cite{mssm} 
\begin{eqnarray}
M_{h,H}^2& = &\frac{1}{2} (M_A^2+M_Z^2+\epsilon) \nonumber\\
 & \times&\Bigg[\,  1 \mp 
\sqrt{1 - 4 \frac{ M_A^2 M_Z^2 \cos^22\beta + \epsilon(M_A^2 
\sin^2\beta + M_Z^2 \cos^2\beta)}{(M_A^2+M_Z^2+\epsilon)^2} }\, 
\Bigg] \ , \\
M_{H^\pm} &=& M_A \left[ 1+ \frac{M_W^2}{M_A^2} \right]^{1/2} \ , \\
\tan 2 \alpha& =& \tan 2 \beta \frac{ M_A^2+M_Z^2} {M_A^2-M_Z^2+ 
\epsilon/\cos{2\beta}} \ ; \ \ -\frac{\pi}{2}<\alpha<0 \ ,
\end{eqnarray}
which include the leading radiative correction
\beq
\epsilon = \frac{3 G_{F}}{\sqrt{2}\pi^2} \frac{m_t^4}{\sin^2\beta}
\log\left( 1+\frac{m_{\tilde{q}}^2}{m_t^2} \right) \ .
\eeq
Here $G_F$ is the Fermi constant and $m_{\tilde{q}}$ is the common
mass scale for the squarks.

The mass terms for the neutral gauginos and Higgsinos originate from the
bilinear Higgs part of the superpotential (the $\mu$ term) and the 
soft--SUSY--breaking gaugino mass terms with masses
$M_2$ and $M_1$ for the SU(2) and U(1) gauginos $\lambda^a$ and
$\lambda^\prime$, respectively. The gaugino mass parameters are
constrained by the GUT relations
\bea
\quad M_1=\frac{5}{3}\tan^2\theta_W M_2\ , 
\quad M_3=\frac{\alpha_s}{\alpha} s^2_W M_2\ .
\label{gut}
\eea
Mixing terms arise from the minimal
coupling terms between Higgs, Higgsino and gaugino fields, where the
Higgs fields have been replaced by their vacuum expectation values.
In terms of two--component spinors the mass terms add up to
\beq
{\cal L}^{\tilde{\chi}^0}_m = -\frac{1}{2}
(-{\rm i}\lambda^\prime,-{\rm i}\lambda^3,\psi^0_{H_1},\psi^0_{H_2}) Y
(-{\rm i}\lambda^\prime,-{\rm i}\lambda^3,\psi^0_{H_1},\psi^0_{H_2})^T + h.c.\ ,
\eeq
with the symmetric mass matrix
\beq
Y = \left(\begin{array}{rcrcrcr}
     M_1 &  & .    &  & .  &  & . \\
     0   &  & M _2 &  & .  &  & . \\
     -M_Z s_W \cos \beta & & M_Z c_W \cos \beta &  &  0   & & . \\
     M_Z s_W \sin \beta  & &-M_Z c_W \sin \beta &  & -\mu & & 0
\end{array}\right)\ .
\eeq
We define the unitary diagonalization matrix $N$ and the matrix
$N^\prime$, which is often more convenient, by the equations
\beq
N^*YN^{-1} = N_{diag},
\ N^\prime = N \cdot \left(\begin{array}{rcrccrr} c_W&&-s_W&&0&&0 \\
                                         s_W&&c_W&&0&&0  \\
                                         0&&0&&1&&0      \\
                                         0&&0&&0&&1
                     \end{array}\right)\ .
\eeq
In general the $\mu$ parameter may be complex and then the
neutralino masses depend also on the phase of $\mu$.
The neutralino mass eigenstates are four
Majorana spinors $\tilde{\chi}_j^0$ given by
\begin{eqnarray}
{\tilde{\chi}}_j^0 & \equiv &
(P_L N_{jk} + P_R N^*_{jk}) \tilde{\Psi}^0_k\ , \\
{\tilde{\Psi}}_k^0 & \equiv &
\left({{-{\rm i}\lambda^\prime}\choose {{\rm i}\bar{\lambda}^\prime}}\ , 
{{-{\rm i}\lambda^3}\choose {{\rm i}\bar{\lambda}^3}}\ ,
{{\psi_{H_1}^0} \choose {\bar{\psi}_{H_1}^0}}\ ,
{{\psi_{H_2}^0} \choose {\bar{\psi}_{H_2}^0}} \right)\ .
\end{eqnarray}

The mass terms for the charged gaugino and Higgsino have the same
origin and, in terms of two-component spinors, read 
\begin{eqnarray}
{\cal L}^{\tilde{\chi}^\pm}_m & = & 
-\frac{1}{2} ( \psi^+\ , \ \ \psi^-)
\left(\begin{array}{cc} 0&X^T \\ X & 0 \end{array}\right)
\left(\begin{array}{r} \psi^+ \\ \psi^- \end{array}\right) , \nonumber\\
X & \equiv & \left(\begin{array}{cc} M_2&M_W \sqrt{2}\sin \beta\\ 
             M_W \sqrt{2} \cos \beta& \mu\end{array}\right)\ ,
\end{eqnarray}
with the abbreviations
\beq
\psi^+_j = (-{\rm i}\lambda^+, \psi_{H_2}^+), \quad \psi^-_j = 
(-{\rm i}\lambda^-, \psi_{H_1}^-)\ .
\eeq
We now define unitary diagonalization matrices $U, V$ by the equation
\beq
U^* X V^{-1} = M_{diag}.
\eeq
The masses of the charginos are given by
\beq
m^2_{\tilde{\chi}^{\pm}_{1,2}}=\frac{1}{2}\Big\{M^2_2+|\mu|^2+2M^2_W\mp
[(M^2_2-|\mu|^2)^2+4M^2_W\cos^22\beta\\
+4M^2_W(M^2_2+|\mu|^2+2M_2{\rm Re}(\mu)\sin 2\beta)]^{1/2}\Big\}.
\eeq

The chargino mass eigenstates are two Dirac spinors $\tilde{\chi}^+_j
$ given by
\begin{eqnarray}
\tilde{\chi}^+_j &\equiv& (P_L V_{jk} + P_R U_{jk}^*)\tilde{\Psi}_k \ ,\\
\tilde{\Psi}_k & \equiv &
 \left({{-{\rm i}\lambda^+} \choose {{\rm i}\bar{\lambda}^-}}\ , 
       {{\psi_{H_2}^+} \choose {\bar{\psi}_{H_1}^-}} \right).
\end{eqnarray}
We abbreviate the charge conjugated fields as
\beq \tilde{\chi}^-_i \equiv \tilde{\chi}^{+ C}_i\ .
\eeq

The mass terms for the scalar quarks originate from the Yukawa couplings
to the Higgs fields (which yield the corresponding quark
masses) and the soft--SUSY--breaking squark mass terms and squark-Higgs
interactions parameterized by $m_{\tilde{q}_{L/R}}$ and $A_q$, 
respectively.
Moreover there are mass and mixing terms from auxiliary field terms
involving one or two Higgs bosons and two squark fields.
The resulting mass matrix for the two squarks of flavor $q$ is:
\beq
{\cal M}^2_{\tilde{q}}=\left(\begin{array}{cc}
       {m^q_L}^2+m_q^2 & m^{q*}_{LR} m_q \\ 
       m^q_{LR} m_q  & {m^q_R}^2+m_q^2 \end{array}\right)\ ,
\eeq
where
\begin{eqnarray}
{m^q_L}^2 & \equiv & m_{\tilde{q}_L}^2 + \cos 2 \beta\ (I^f_3-Q_fs_W^2) M_Z^2\ , 
\label{b28}\\
{m^q_R}^2 & \equiv & m_{\tilde{q}_R}^2 + \cos 2 \beta\ (Q_fs_W^2) M_Z^2
\end{eqnarray}
and  
\beq
m^q_{LR} \equiv A_q - \mu^*\{\cot\beta,\tan\beta\}\ ,
\label{mlr}
\eeq
for $\{$up, down$\}$--type squarks, respectively.

This hermitian mass matrix is diagonalized by a unitary matrix $S^q$, so
we can write the squark mass eigenstates of flavor $q$ as
\begin{eqnarray}
\tilde{q}_{1} & = & S^q_{11}\ \tilde{q}_{L} + S^q_{12}\ 
\tilde{q}_{R}\ ,\nonumber\\
\tilde{q}_{2} & = & S^q_{21}\ \tilde{q}_{L} + S^q_{22}\ 
\tilde{q}_{R}\ .
\end{eqnarray}
In general, $m^q_{LR} = |m^q_{LR}|e^{{\rm i}\varphi_{\tilde q}}$ is a complex 
number and $S^q$ may be written as
\beq S^{q} = 
 \left(\begin{array}{rr} 
  e^{{\rm i}\varphi_{\tilde q}/2}\cos\theta_{\tilde q} &  
  e^{-{\rm i}\varphi_{\tilde q}/2}\sin\theta_{\tilde q} \\
 -e^{{\rm i}\varphi_{\tilde q}/2}\sin\theta_{\tilde q} &  
  e^{-{\rm i}\varphi_{\tilde q}/2}\cos\theta_{\tilde q}
                        \end{array}\right)\ .
\eeq

The mass terms of the sleptons arise in the same way as the
squark masses. The main difference appears for the sneutrinos: there
is only one sneutrino for every generation, $\tilde{\nu}_l$, and hence there
is no mixing. Moreover, we cannot add trilinear soft--breaking terms
to shift the masses of the sneutrinos, whose value is given just by
$m^2_{\tilde{\nu}}=m^2_{\tilde{l}_L}+1/2\cos2\beta M^2_Z$. (See
Eq.~(\ref{b28})).

In general, the physical sfermion masses are given by
\beq
m^2_{\tilde{f}_{1,2}}=m^2_f+\frac{1}{2}\left\{({m^f_R}^2+{m^f_L}^2)\mp
\sqrt{({m^f_R}^2-{m^f_L}^2)^2+4m^2_f|m^f_{LR}|^2}\right\},
\eeq
independent of the phase of $m^f_{LR}$.

\subsection{Physical Phases in the MSSM with complex couplings}

As we do not constrain the parameters of the MSSM to be real, the
following parameters may take complex values: the
Yukawa couplings, $\mu$, the soft parameters $m_{12}^2$, $M_1$, $M_2$,
$M_3$ and the $A$ parameters.
But not all combinations of phases in these parameters lead to
different physical results because several phases can be absorbed by
redefinitions of the fields. We now describe a procedure to eliminate
the unphysical phases. We assume the GUT relation between the $M_i$,
so they have one common phase. The only remaining phases will be
chosen to be those of $\mu$, the $A$ parameters and, for three
generations, one phase for all the Yukawa couplings (the $\delta_{\rm CKM}$).

Analogous to the Standard Model case, the Yukawa couplings can be
changed by redefinitions of the quark superfields in such a way that
there remains only one phase for three generations and only real
couplings for less than three generations.

\begin{table}[htb]
\caption{\em The charges $n_i$ for three U(1) symmetries that leave
  ${\cal L}_{\rm MSSM}$ invariant.}
\label{PQCharges}
\vspace{0.3cm}
\center
\begin{tabular}{|c|r|r|r|r|r|r|r|r|r|c|}
\hline
U(1) & $M_i$ & $A$ & $m_{12}^2$ & $\mu$\ & $H_1$ & $H_2$ & $QU$\ 
     & $QD$\ & $LE$\ & $\theta$\\
\hline
\hline 
$PQ$  & 0 & 0 & $-1$ & $-1$ & 1/2 & 1/2 & $-1/2$ & $-1/2$ 
      & $-1/2$ & 0 \\
$R_1$ & $-1$ & $-1$ & 0 & 1 & 0 & 0 & 1 & 1 & 1 & $-1/2$ \\
\hline 
\end{tabular}
\vspace*{0.3cm}
\end{table}

Following \cite{relax} we consider two U(1) transformations $PQ$
and $R_1$ that do not only transform the fields but also the
parameters of the MSSM. In table \ref{PQCharges} the charges $n_i$ of
the various quantities are displayed with which the Lagrangian ${\cal
  L}_{\rm MSSM}$ is invariant under the combined multiplications with
$e^{i\alpha n_i}$. The first transformation is a Peccei-Quinn
symmetry and $R_1$ is an R symmetry that also transforms the $\theta$
variables appearing in the arguments of the superfields in ${\cal
  L}_{\rm MSSM}$. 

Since ${\cal L}_{\rm MSSM}$ is invariant under these transformations,
so are the physical predictions of the MSSM. If the parameters $M_i$, $A$,
$m_{12}^2$ and $\mu$  have the phases $\varphi_M$, $\varphi_A$,
$\varphi_{m_{12}^2}$ and $\varphi_\mu$, we first apply $R_1$ with the
angle $\varphi_M$, then $PQ$ with the angle $\varphi_{m_{12}^2}$ and
obtain for an arbitrary observable:
\beq
\begin{array}{clcccc}
&\sigma(|\mu|,|A|,|M_i|,|m_{12}^2|,&\varphi_\mu,&\varphi_A,
&\varphi_M,&\varphi_{m_{12}^2})
\\ = & 
\sigma(|\mu|,|A|,|M_i|,|m_{12}^2|,&\varphi_\mu+\varphi_M,
&\varphi_A-\varphi_M,&0,&\varphi_{m_{12}^2})
\\ = & 
\sigma(|\mu|,|A|,|M_i|,|m_{12}^2|,&\varphi_\mu+\varphi_M-\varphi_{m_{12}^2},
&\varphi_A-\varphi_M,&0,&0).
\end{array}
\eeq
So the physical predictions only depend on the absolute
values of the parameters and the phases
\beq
\phi_A\equiv {\rm arg}(A M_i^*), \quad \phi_B\equiv {\rm arg}(\mu A m_{12}^{2*})
\ .
\eeq
One can replace these phases by another set where only the $\mu$ and the $A$ 
parameters are complex. In our choice, the phase of $A$ is traded for the phase 
$\varphi_{\tilde f}$ of the off--diagonal term in the corresponding sfermion 
mixing matrix: $m^f_{LR}\equiv A_f-\mu^*\{\cot,\tan\}\beta$ (\ref{mlr}). 
Relaxing universality for the soft--breaking terms, every $A_f$ contains a 
different CP--violating phase. 

\subsection{Vertex factors}

\begin{figure}[htb]
\begin{center}
\begin{tabular}{c}
\epsfig{figure=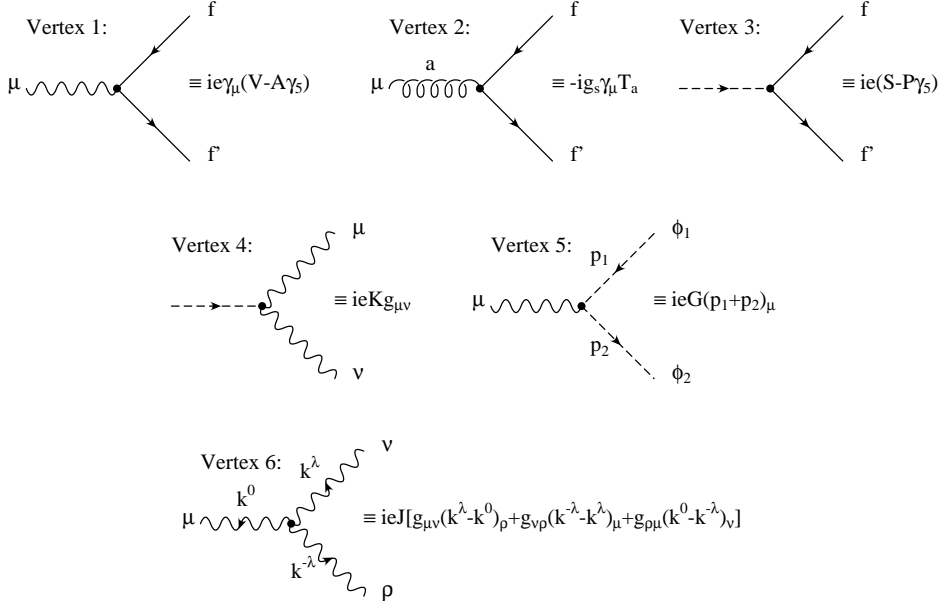,width=0.9\linewidth}
\end{tabular}
\end{center}
\caption{\em Generic Vertices.\label{fig:ver}}
\vspace*{0.3cm}
\end{figure}

We employ the notation and conventions of the previous sections. 
The list of generic vertices is shown in Fig.~\ref{fig:ver}.

    \subsubsection{Couplings in the SM}

\begin{description}

\item{\bf{Vertex 1:}} Coupling of electroweak gauge bosons and 
fermions: ${\rm i}e\gamma_\mu (V-A\gamma_5)$.
\beq
\gamma \bar{f}f: & V&=-Q_f \ , \ \ A=0 \ ,  \\
Z\bar{f}f:       & V&=-\frac{v_f}{2s_Wc_W} \ ,  \ \ A=-\frac{a_f}{2s_Wc_W} \ ,
\\
W^\pm \bar{f}'f: & V&=A=-\frac{1}{2\sqrt{2}s_W}\ ,
\eeq 
with $v_f\equiv(I^f_3-2s^2_WQ_f)$, $a_f\equiv I^f_3$.

\item{\bf{Vertex 2:}} Coupling of gluons and quarks: 
$-{\rm i}eg_s\gamma_\mu T_a$.

\item{\bf{Vertex 3:}} Coupling of two fermions and one Higgs boson
(${\cal H}\bar{f}'f$): ${\rm i}e (S-P\gamma_5)$.
\beq
H^0\bar{f}f:          & S=-\mu_f \ ,  & P=0 \ , \\
\chi \bar{f}f:        & S=0 \ ,       & P=-2{\rm i}I_3^f\mu_f \ , \\
\phi^\pm \bar{f}'f:   & S= \sqrt{2}I_3^f[\mu_f-\mu_{f'}] \ ,  
                      & P=-\sqrt{2}I_3^f[\mu_f+\mu_{f'}]\ ,
\eeq
with $\mu_f\equiv\displaystyle\frac{m_f}{2s_WM_W}=
\displaystyle\frac{m_f}{2s_Wc_WM_Z}$.

\item{\bf{Vertex 4:}} Coupling of one Higgs boson and two gauge 
bosons: ${\rm i}e K g_{\mu\nu}$.
\beq
H^0ZZ:                  & K=&\displaystyle\frac{M_Z}{s_Wc_W} \ , \\
H^0W^\pm W^\mp:         & K=&\displaystyle\frac{M_W}{s_W} \ , \\
\phi^\pm W^\mp\gamma:   & K=&M_W \ ,  \\
\phi^\pm W^\mp Z:       & K=&-M_Zs_W \ .  
\eeq 
The rest of the couplings are $K=0$.

\item{\bf{Vertex 5:}} Coupling of one gauge boson and 
two Higgs bosons ($V\phi_1\phi_2$): \\${\rm i}e G (p_1+p_2)_\mu$.

\beq
\gamma\phi^\lambda\phi^{-\lambda}:    &G=&-\lambda \ , \\
Z\phi^\lambda\phi^{-\lambda}:&G=&-\lambda\displaystyle\frac{\cos2\theta_W}
                             {2s_Wc_W}    \ , \\
Z\chi H^0:                   &G=&\displaystyle\frac{{\rm i}}{2s_Wc_W} \ , \\
W^\lambda\phi^{-\lambda}H^0: &G=&\displaystyle\frac{\lambda}{2s_W} \ , \\
W^\lambda\phi^{-\lambda}\chi:&G=&\displaystyle\frac{{\rm i}}{2s_W} \ .
\eeq 
Interchanging the two Higgs bosons causes the coupling constant to change sign.

\item{\bf{Vertex 6:}} Coupling of three gauge bosons 
(outgoing momenta): 
$${\rm i}e J \{g_{\mu\nu}(k^\lambda-k^0)_\rho
       +g_{\nu\rho}(k^{-\lambda}-k^\lambda)_\mu
       +g_{\rho\mu}(k^0-k^{-\lambda})_\nu \} \ .$$
\beq
\gamma W^\lambda W^{-\lambda}: &J=&-\lambda \ , \\
Z W^\lambda W^{-\lambda}:      &J=&-\lambda\displaystyle\frac{c_W}{s_W} \ .
\eeq 

\end{description}

    \subsubsection{Couplings in the MSSM}

\begin{description}

\item{\bf{Vertex 1:}} Coupling of two neutralinos or charginos and
one gauge boson: ${\rm i}e\gamma_\mu (V-A\gamma_5)$.

The fermion flow direction in our Feynman graphs is fixed by the
outgoing fermion antifermion pair. The relation between the gauge
boson vertex factors for the two different fermion flow directions is
obtained by substituting spinors by charge conjugated spinors in the
interaction term:
\beq
V_\mu \bar{\tilde{\chi}}^+ \gamma^\mu (P_L g_L + P_R g_R)
\tilde{\chi}^+
=
V_\mu \bar{\tilde{\chi}}^- \gamma^\mu (-P_L g_R - P_R g_L)
\tilde{\chi}^-\ . 
\eeq
Taking into account also the symmetry factor $S=2$ for the
neutralino coupling and $S=1$ for the charginos, the boson vertex
factors $V\equiv(g_L+g_R)/2$ and $A\equiv(g_L-g_R)/2$ are given by:
\beq
Z\bar{\tilde{\chi}}^0_j\tilde{\chi}^0_k: & 
g_L=&\displaystyle\frac{1}{2s_Wc_W}
(N^\prime_{k4}N^{\prime *}_{j4}-N^\prime_{k3}N^{\prime *}_{j3})\ , \\ &  
g_R=&\displaystyle\frac{1}{2s_Wc_W}
(N^{\prime *}_{k3}N^\prime_{j3}-N^{\prime *}_{k4}N^\prime_{j4})\ , \\ 
Z\bar{\tilde{\chi}}^+_k {\tilde{\chi}}^+_j
: & 
g_L=&\displaystyle - \frac{1}{s_Wc_W}
\left[\left(\frac{1}{2} - s_W^2\right) V^*_{k2}V_{j2}+c_W^2 
V^*_{k1}V_{j1}\right]
\ , \\ &
g_R=&\displaystyle - \frac{1}{s_Wc_W}
\left[\left(\frac{1}{2} - s_W^2\right) U_{k2}U^*_{j2}+c_W^2 
U_{k1}U^*_{j1}\right]
\ , \\  
Z\bar{\tilde{\chi}}^-_j\tilde{\chi}^-_k: & 
g_L=&\displaystyle\frac{1}{s_Wc_W}
\left[\left(\frac{1}{2} - s_W^2\right) U_{k2}U^*_{j2}+c_W^2 
U_{k1}U^*_{j1}\right]
\ , \\ & 
g_R=&\displaystyle\frac{1}{s_Wc_W}
\left[\left(\frac{1}{2} - s_W^2\right) V^*_{k2}V_{j2}+c_W^2 
V^*_{k1}V_{j1}\right]\ ,
\eeq
and
\beq
\gamma\bar{\tilde{\chi}}^+_k {\tilde{\chi}}^+_j
: & 
g_L=g_R=&-\delta_{jk}\ , \\  
\gamma\bar{\tilde{\chi}}^-_j\tilde{\chi}^-_k: & 
g_L=g_R=&\delta_{jk}\ .
\eeq

\item{\bf{Vertex 2:}} There is no genuine supersymmetric vertex of
this kind.

\item{\bf{Vertex 3:}} There are three couplings of the kind
${\rm i}e (S-P\gamma_5)$ :

(Some of the couplings are more easily written in terms of
$g_{L,R}\equiv S\pm P$).

\noindent\hspace{-0.33cm}$\bullet$  
Coupling of one neutralino or chargino to fermions and scalar fermions. 

The couplings of neutralinos and charginos to quarks and scalar quarks are
given by
\begin{eqnarray}
{\cal L}_{\tilde{\chi} q \tilde{q}} & = & e\tilde{q}^\dagger
\bar{\tilde{\chi}}(P_L g_L + P_R g_R) q 
+ e\bar{q} (P_R g_L^* + P_L g_R^*) \tilde{\chi} \tilde{q}\ .
%
\end{eqnarray}


\underline{down-type quarks} ($i=1,3,5$):
\begin{eqnarray}
\tilde{q}_{i,k}^\dagger \bar{\tilde{\chi}}^0_j q_i :&
g_L=& 
-\sqrt{2} \left[
      \left(Q_i  N^{\prime *}_{j1}
      + (I^i_3-Q_i s_W^2)  \frac{1}{s_Wc_W} N^{\prime *}_{j2} \right)
      S_{k1}^i  \right.\nonumber\\
&&\left.
      + \frac{ m_{q_i}}{2 M_W s_W\cos\beta}N^{\prime *}_{j3}
      S_{k2}^i
      \right]\ , 
\\
&g_R=& 
-\sqrt{2} \left[
      - \left(Q_i     N^\prime_{j1}
      + (-Q_i s_W^2)  \frac{1}{s_Wc_W} N^\prime_{j2} \right)
      S_{k2}^i \right.\nonumber\\
&&\left.
     + \frac{ m_{q_i}}{2 M_W s_W\cos\beta} N^\prime_{j3}
     S_{k1}^i \right]\ ,
\\
\tilde{q}_{(i+1),k}^\dagger \bar{\tilde{\chi}}^-_j q_i :&
g_L=& 
\frac{1}{s_W} \left[  - V^*_{j1} S_{k1}^{i+1}
          + \frac{m_{q_{i+1}}}{\sqrt{2} M_W \sin\beta}
            V^*_{j2} S_{k2}^{i+1} \right]\ ,
\\
&g_R=& 
\frac{1}{s_W} \left[   \frac{m_{q_i}}{\sqrt{2} M_W \cos\beta}
           U_{j2}  S_{k1}^{i+1}
\right]\ .
\end{eqnarray}

\underline{up-type quarks} ($i=2,4,6$):
\begin{eqnarray}
\tilde{q}_{i,k}^\dagger \bar{\tilde{\chi}}^0_j q_i :&
g_L=& 
-\sqrt{2} \left[
       \left(Q_i   N^{\prime *}_{j1}
       + (I^i_3 - Q_i s_W^2)  \frac{1}{s_Wc_W} N^{\prime *}_{j2} \right)
       S_{k1}^i  \right.\nonumber\\
&&\left.
       + \frac{ m_{q_i}}{2 M_W s_W\sin\beta} N^{\prime *}_{j4}
       S_{k2}^i \right]\ ,
\\
&g_R=& 
-\sqrt{2} \left[
      - \left(Q_i   N^\prime_{j1}
         + (-Q_i s_W^2)  \frac{1}{s_Wc_W} N^\prime_{j2} \right)
         S_{k2}^i \right.\nonumber\\
&&\left.
         + \frac{m_{q_i}}{2 M_W s_W\sin\beta} N^\prime_{j4}
         S_{k1}^i \right]\ ,
\\
\tilde{q}_{(i-1),k}^\dagger \bar{\tilde{\chi}}^+_j q_i :&
g_L=& 
\frac{1}{s_W} \left[ - U^*_{j1} S_{k1}^{i-1}
         + \frac{m_{q_{i-1}}}{\sqrt{2} M_W \cos\beta}
           U^*_{j2} S_{k2}^{i-1} \right]\ ,
\\
&g_R=& 
\frac{1}{s_W} \left[ \frac{m_{q_i}}{\sqrt{2} M_W \sin\beta}
           V_{j2} S_{k1}^{i-1} \right]\ .
\end{eqnarray}

The couplings of neutralinos and charginos to leptons and scalar leptons are
given analogously, performing the following substitutions:
\beq
i=1,3,5&:&\tilde{q}_{i1}\to\tilde{l}_1
\ , \ \ S^i_{jk}\to S^l_{jk}\ , \ \ Q_i=-1\ , \ \ I^i_3=-\frac{1}{2}\ ,
\ \ m_{q_i}=m_l \ ,\nonumber\\
 & & \tilde{q}_{i2}\to\tilde{l}_2\ ,\nonumber\\
i=2,4,6&:&\tilde{q}_{i1}\to\tilde{\nu}_l
\ , \ \ S^i_{jk}\to \delta_{jk}\ , \ \ Q_i=0\ , \ \ I^i_3=\frac{1}{2}\ , 
\ \ m_{q_i}=0 \ ,\nonumber\\
 & & \tilde{q}_{i2}\ \mbox{does not exist} \ . \nonumber
\eeq

\noindent\hspace{-0.33cm}$\bullet$  
Coupling of one gluino to a quark and a scalar quark.

The interaction between gluinos, quarks and squarks is described by
the terms
\begin{eqnarray}
{\cal L} & = & \tilde{q}^\dagger
e\overline{\tilde{g}}^a (P_L g_L + P_R g_R) \frac{\lambda^a}{2} q 
+ e\overline{q} (P_R g_L^* + P_L g_R^*) \frac{\lambda^a}{2}
\tilde{g}^a \tilde{q}\ ,
\end{eqnarray}
yielding the vertex factors
\beq
{\rm i}e(P_L g_L + P_R g_R)\frac{\lambda^a}{2},\quad
{\rm i}e(P_R g_L^* + P_L g_R^*)\frac{\lambda^a}{2}.
\eeq
In our calculations the Gell--Mann matrices appear only in the
combination 
\beq
\sum_{a=1}^8 \left(\frac{\lambda^a \lambda^a}{4} \right)_{AB} =
C_2(F)\delta_{AB} = \frac{4}{3}\delta_{AB} .
\eeq
The couplings are
\begin{eqnarray}
\tilde{q}_{i}^\dagger \overline{\tilde{g}} q 
:
&eg_L =& -\sqrt{2} \displaystyle{g_s}  S^q_{i1} \ , \\
&eg_R =& +\sqrt{2} \displaystyle{g_s}  S^q_{i2} \ .
\end{eqnarray}

\noindent\hspace{-0.33cm}$\bullet$  
Coupling of two fermions and one Higgs boson (${\cal H}\bar{f}'f$).
\begin{eqnarray}
H\bar{u}u: & S=
-\mu_u\sin\alpha/\sin\beta\ , & P=0 \ , \label{c78} 
\\
H\bar{d}d: & S=
-\mu_d\cos\alpha/\cos\beta\ , & P=0\ , 
\\
h\bar{u}u: & S=
-\mu_u\cos\alpha/\sin\beta\ , & P=0\ , 
\\
h\bar{d}d: & S=
\mu_d\sin\alpha/\cos\beta\ , & P=0\ , 
\\
A\bar{u}u: &S=0 \ , & 
P=-{\rm i}\mu_u\cot\beta\ ,  
\\
A\bar{d}d: &S=0  \ , & 
P=-{\rm i}\mu_d\tan\beta \ , \label{c83}
\\
H^+\bar{u}d: & S=
\displaystyle\frac{1}{\sqrt{2}}(\mu_u\cot\beta +\mu_d\tan\beta) \ , &
\nonumber\\
&
P=\frac{1}{\sqrt{2}}(\mu_u\cot\beta-\mu_d\tan\beta) \ , &
\\
G^0\bar{f}f: & S=0 \ ,        & P=-2{\rm i}I_3^f\mu_f \ , \label{c85}\\
G^\pm \bar{f}'f: & S= \sqrt{2}I_3^f[\mu_f-\mu_{f'}] \ ,  
           & P=-\sqrt{2}I_3^f[\mu_f+\mu_{f'}]\ .
\end{eqnarray}
with $\mu_f\equiv m_f/2s_WM_W=m_f/2s_Wc_WM_Z$. For the vertices corresponding 
to the hermitian conjugated fields we have to replace ($S,P$) by ($S^*,-P^*$).

\item{\bf{Vertex 4:}} Coupling of one Higgs boson and two gauge bosons:
${\rm i}e K g_{\mu\nu}$.

The couplings of a neutral Higgs to two $Z$ bosons are:
\begin{eqnarray}
hZZ: & K=&\frac{M_Z}{s_Wc_W} \sin(\beta-\alpha)\ , \\
HZZ: & K=&\frac{M_Z}{s_Wc_W} \cos(\beta-\alpha)\ . 
\end{eqnarray}
And the only nonzero couplings of a charged Higgs to neutral gauge and $W$ 
bosons are:
\begin{eqnarray}
G^\pm W^\mp \gamma:     & K=& M_W \ , \\
G^\pm W^\mp Z:          & K=& -M_Z s_W \ .
\end{eqnarray}
The rest of the couplings are $K=0$.

\item{\bf{Vertex 5:}} There are two types of couplings $Z\phi_1\phi_2$
of the kind ${\rm i}e G (p_1+p_2)_\mu$\ . 

\noindent\hspace{-0.33cm}$\bullet$  
Coupling of two Higgs bosons and one neutral gauge boson.

\begin{eqnarray}
ZAH: &G=&
-\frac{{\rm i}\sin(\beta-\alpha)}{2 s_Wc_W} \ ,
\\
ZAh: &G=&
\frac{{\rm i}\cos(\beta-\alpha)}{2 s_Wc_W} \ ,
\\
ZG^0H: &G=&
\frac{{\rm i}\cos(\beta-\alpha)}{2 s_Wc_W} \ ,
\\
ZG^0h: &G=&
\frac{{\rm i}\sin(\beta-\alpha)}{2 s_Wc_W} \ ,
\\
ZH^\lambda H^{-\lambda}: &G=&
-\lambda \frac{\cos2\theta_W}{2s_Wc_W} \ ,  
\\
Z G^\lambda G^{-\lambda}: &G=&
-\lambda \frac{\cos2\theta_W}{2s_Wc_W} \ ,
\\
\gamma H^\lambda H^{-\lambda}: &G=&
-\lambda  \ ,  
\\
\gamma G^\lambda G^{-\lambda}: &G=&
-\lambda  \ .    
\end{eqnarray}
Interchanging the $\phi_1$ and $\phi_2$ causes the coupling 
constant to change sign.

\noindent\hspace{-0.33cm}$\bullet$  
Coupling of two scalar fermions and one neutral gauge boson.
\beq
\gamma\tilde{q}_{i,j}\tilde{q}^\dagger_{i,k}: & G= &
-Q_i \delta_{jk} \ , 
\\
Z\tilde{q}_{i,j}\tilde{q}^\dagger_{i,k}: & G= &
-\frac{1}{s_Wc_W} 
\left[(I^i_3 - Q_i s_W^2) S^{i*}_{j1} S^{i}_{k1}
- Q_i s_W^2 S^{i*}_{j2} S^{i}_{k2} \right] \ .
\eeq

\item{\bf{Vertex 6:}} There is no genuine supersymmetric vertex of
this kind.

\end{description}


\end{document}